\documentclass[aps,prx,twocolumn,groupedaddress]{revtex4-2}
\usepackage[utf8]{inputenc}
\usepackage[T1]{fontenc}
\usepackage{graphics}
\graphicspath{{figures/}}
\usepackage{xcolor}
\usepackage{mathtools}
\usepackage{braket}
\usepackage{amssymb}
\usepackage{bm}
\usepackage{tikz}

\usepackage{hyperref}
\hypersetup{colorlinks=true,allcolors=blue}

\newcommand*{\s}[2]{S^{#1}_{#2}}
\newcommand*{\vs}[1]{{\mathbf S}_{#1}}
\newcommand*{\up}{\uparrow}
\newcommand*{\down}{\downarrow}
\newcommand*{\deltaeff}{\tilde\delta}
\newcommand*{\Jeff}{\tilde J}
\definecolor{darkred}{RGB}{204, 0, 0}

\DeclareMathOperator{\re}{Re}
\DeclareMathOperator{\sign}{sign}
\DeclarePairedDelimiter{\norm}{\lVert}{\rVert}
\DeclarePairedDelimiter{\mean}{\langle}{\rangle}
\DeclarePairedDelimiter{\ceil}{\lceil}{\rceil}
\DeclarePairedDelimiter{\abs}{\lvert}{\rvert}

\begin{document}


\title{Spin many-body phases in standard and topological waveguide QED simulators}


\author{M. Bello}
\email[]{miguel.bello@mpq.mpg.de}
\affiliation{Max-Planck-Institut f\"ur Quantenoptik, Hans-Kopfermann-Strasse 1, 85748 Garching, Germany.}
\author{G. Platero}
\affiliation{Instituto de Ciencia de Materiales de Madrid, CSIC, 28049 Madrid, Spain.}
\author{A. Gonz\'alez-Tudela}
\email[]{a.gonzalez.tudela@csic.es}
\affiliation{Instituto de Física Fundamental, IFF-CSIC, Calle Serrano 113b, Madrid 28006, Spain.}


\date{\today}

\begin{abstract}
Waveguide QED has emerged as a powerful analog quantum simulator due to the possibility of mediating versatile spin-spin interactions with tunable sign, range, and even chirality. Yet, despite their potential, the many-body phases emerging from these systems have only been scarcely explored. Here, we characterize the many-body phases of a large class of spin models that can be obtained in such waveguide QED simulators, and uncover, importantly, the existence of symmetry-protected topological phases with large-period magnetic orderings with no analogue in other state-of-art simulators. We explain that these phases emerge from the unique combination of long-range and dimerized interactions appearing in these platforms, and propose several experimental observables to characterize them. Finally, we also develop an adiabatic protocol to prepare such states, and analyze its performance with the main decoherence source of these systems.
\end{abstract}


\maketitle

\section{Introduction \label{sec:intro}}

The importance of quantum spin problems in several areas of physics and computation is hard to overestimate~\cite{auerbachbook94a}. For example, interacting spin systems are believed to be instrumental for understanding the behaviour of high-Tc superconductors~\cite{lee06a} or exciton transport~\cite{Clegg2006}, they are paradigmatic models of quantum phase transitions~\cite{Sachdev2011}, and, beyond physics, their ground states can codify the solution of many NP-hard optimization problems~\cite{Lucas2014}. Irrespective of the context where they appear, understanding their many-body behaviour represents an outstanding computational challenge due to the exponential growth of their Hilbert space with the system size. This difficulty becomes critical for frustrated spins systems~\cite{Henelius2000}, where the classical algorithms that avoid the exponential growth of resources required, like Montecarlo, do not perform well. An alternative way to study these problems consists in building analogue quantum simulators---well-controlled systems that replicate the interactions of the models to be understood~\cite{cirac12a}. In this way, by preparing, manipulating, and measuring the simulator, one can get information about the spin model that would be difficult to obtain otherwise.

Among the different platforms available, atomic systems in their different incarnations are nowadays the leading technology for analogue simulation of quantum magnetism. For example, the internal levels of trapped ion chains can behave as effective spin systems that interact through their collective motion~\cite{porras04a}, with interactions decaying generally as $1/r^3$ with the distance $r$ between ions~\cite{nevado16a}. These simulators have been experimentally realized by many groups already~\cite{friedenauer08a,kim09a,kim10a,Edwards2010,Islam2011,Lin2011SharpSpins}, who have used them, for example, to probe the phase diagram of interacting spin chains \cite{Edwards2010}. Another implementation that can be used to simulate quantum spin models are neutral atoms trapped in optical lattices. To do so, one can define a pseudo-spin degree of freedom either by using the occupation difference between neighboring lattice sites~\cite{Sachdev2002,simon11a,ma11a,Meinert2013,Meinert2014} or the hyperfine levels of the atoms~\cite{duan03a,Garcia-Ripoll2003,Kuklov2003,trotzky08a,Nascimbene2012,fukuhara13a,fukuhara13b,Hild2014}, leading to ZZ or Heisenberg type interactions, respectively. Unfortunately, irrespective of the method chosen, the interaction range is restricted to nearest neighbours, limiting the frustration that can be obtained in these systems. To extend this range, one can excite the atom to a Rydberg state, yielding $1/r^6$ (ZZ) or $1/r^3$ (XX) interactions depending on the parity of the Rydberg state~\cite{Weimer2010,Lesanovsky2011,labuhn16a,bernien17a,DeLeseleuc2019,Browaeys2020}. This increase in the interaction range comes at the price of introducing additional noise into the system due to the Rydberg state lifetime, although it has still enabled the observation of novel frustrated spin many body phases experimentally~\cite{bernien17a} coming from the longer-range nature of the interactions.

Despite the unquestionable success of all these platforms, the models they can simulate are limited by the interactions that appear naturally in these systems, which generally have a fixed range, sign, or type. Waveguide QED setups, where quantum emitters couple to one-dimensional propagating fields~\cite{vetsch10a,goban13a,thompson13a,beguin14a,goban15a,hood16a,solano16a,Kim2019,luan2020,Samutpraphoot2020,laucht12a,lodahl15a,evans18a,vanloo13a,liu17a,mirhosseini18a,Mirhosseini2019,kim2021,krinner18a}, have been suggested as an alternative powerful paradigm to study quantum magnetism circumventing these tunability limitations~\cite{douglas15a,Gonzalez-Tudela2015b,Hung2016,Chang2018,Bello2019,leonforte2021,saxena2021photonic}. The key idea is that when the emitter's transition frequencies are tuned to a photonic bandgap, the photons localize around them forming bound-states~\cite{bykov75a,john90a,kurizki90a} that can mediate coherent spin interactions between the emitters. 
The sign, range, and even chirality of these bound-state-mediated interactions can be largely tuned by either properly engineering the waveguide or through a Raman-assisted process~\cite{douglas15a,Gonzalez-Tudela2015b,Hung2016,Chang2018,Bello2019,leonforte2021}. Besides, by combining several Raman-assisted processes one can further tune the nature of the interactions from the naturally appearing exchange interactions to Ising-type \cite{douglas15a} and combinations of them~\cite{Gonzalez-Tudela2015b,Hung2016}. These exciting perspectives have triggered many experimental advances in the field and nowadays we count with many platforms where natural~\cite{goban13a,thompson13a,hood16a,Kim2019,luan2020,Samutpraphoot2020} and artificial atoms~\cite{laucht12a,lodahl15a,evans18a}, superconducting qubits~\cite{liu17a,mirhosseini18a,Mirhosseini2019,kim2021}, or even simulated emitters~\cite{krinner18a}, couple to waveguides with photonic bandgaps. Remarkably, despite their great potential for quantum simulation, the expected many-body phases that 
emerge in these systems have only been scarcely studied~\cite{manzoni2017manybody,li2020random,Bello2019}.

In this manuscript, we cover this gap providing the first complete analysis of the many-body spin phases that appear in a large variety of waveguide QED setups. Based on previous works~\cite{douglas15a,Gonzalez-Tudela2015b,Hung2016,Chang2018,Bello2019,leonforte2021}, we consider a generalized XXZ spin model with interactions that originate in a photonic SSH waveguide. This kind of waveguide can produce long-range, dimerized (chiral) interactions with staggered sign, and it describes both the trivial and topological waveguide cases.
Using density matrix renormalization group techniques~\cite{schollwock11a,Hauschild2018} and exact diagonalization, we find and characterize frustrated many-body phases with long-range magnetic order emerging from the competition of all these features, and propose observables to characterize them experimentally. Importantly, we show that some of these phases are symmetry-protected topological phases displaying quantized many-body Berry phases~\cite{hatsugai2006,hirano2008,kariyado2018}. Our results thus demonstrate the potential of waveguide QED platforms as an analogue quantum simulator of frustrated spin models, including those considering the interplay between topology and long-range interactions, which have aroused lately a lot of interest~\cite{nevado17a,gong16a,patrick17a,Bermudez2017Long-rangeIons}.

The manuscript is structured as follows: in Section~\ref{sec:model} we describe the model that we will consider, explaining the shape of the waveguide-mediated interactions in different bandgap configurations. In Section~\ref{sec:standard} we study the many-body phases that emerge in conventional (undimerized) waveguides, where both long-range and/or staggered-sign interactions appear depending on the type of bandgap chosen. Then, in Section~\ref{sec:topological} we consider the case of topological-waveguides~\cite{Bello2019,leonforte2021,kim2021}, where, in addition, the dimerization and chirality of the interactions can be tuned by modifying the properties of the waveguide. In Section~\ref{sec:adiabatic}, we design and benchmark an adiabatic protocol to prepare the most relevant phases, and study its resilience to one of the most relevant decoherence sources, that is, emitter excitation loss. Finally, we summarize our findings in Section~\ref{sec:conclusions}.

\section{Generalized spin models in waveguide QED \label{sec:model}}

\begin{figure}[!tb]
  \includegraphics{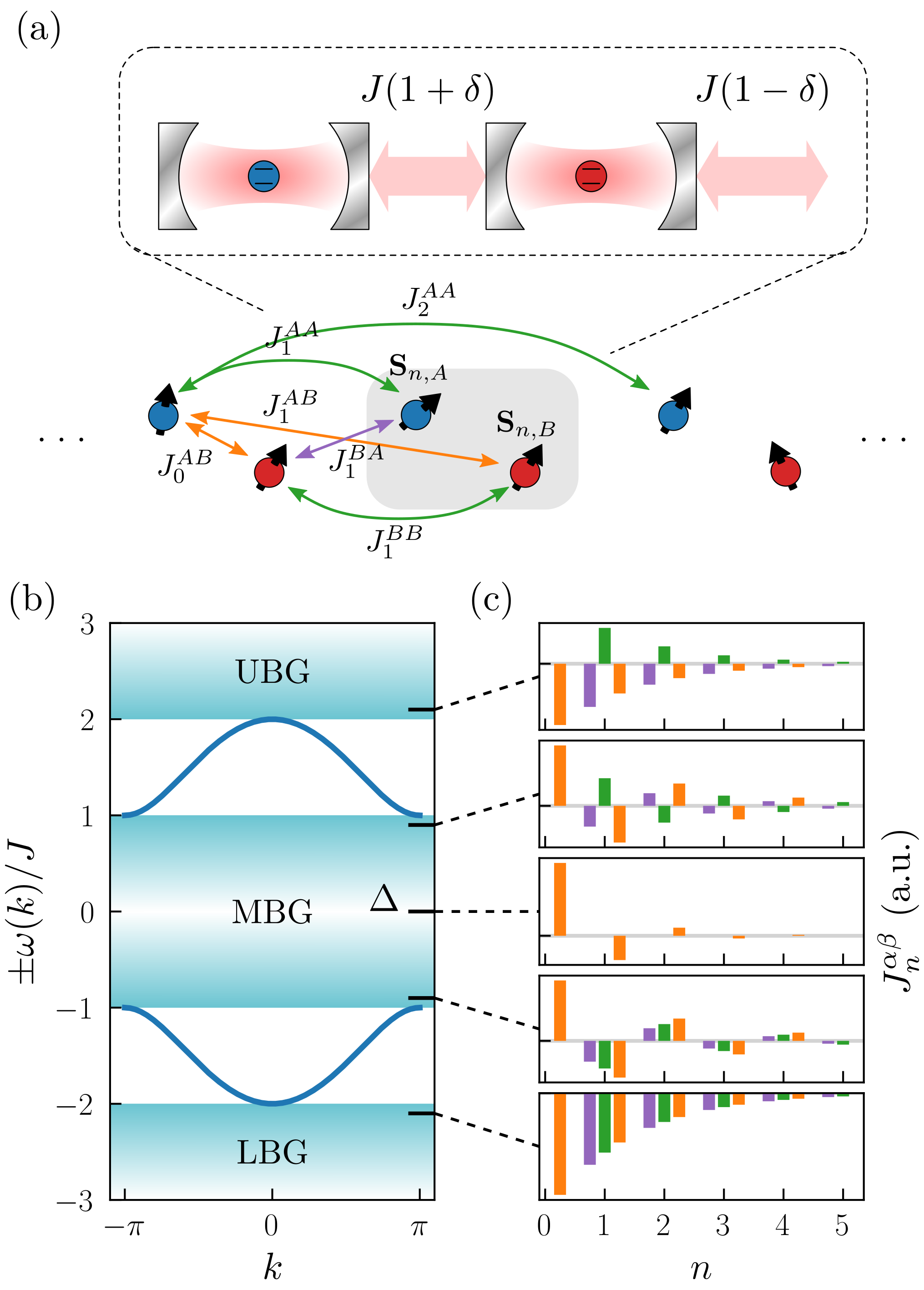}
  \caption{\textbf{Waveguide-mediated interactions.} (a) Schematic picture of the system under consideration. The blue (red) dots represent spins in 
  the $A$ ($B$) sublattice and the colored arrows represent couplings 
  between different spins. 
  (b) Energy bands of the photonic SSH model for $\delta=0.5$. The shaded
  areas mark the upper, middle, and lower bandgaps.
  (c) Interaction constants as a function of distance for different values 
  of the emitter transition frequency $\Delta$, as indicated by horizontal 
  segments on the right side of plot (b). Different spin interactions are 
  colored as in the system diagram (a). Positive (negative) coupling 
  constants are plotted above (below) the grey horizontal line.
  \label{fig:schematics}}
\end{figure}

The system we will study consists of $N$ quantum emitters that interact locally with the photonic analogue of the SSH model~\cite{Bello2019,leonforte2021}. This waveguide can be described as a set of tunnel-coupled cavities with alternating hopping amplitudes $J(1\pm \delta)$, all with the same resonant frequency $\omega_a$, see Fig.~\ref{fig:schematics}(a). This model covers both the cases of conventional waveguides, if $\delta=0$, where a single energy band with a simple dispersion appears, $\omega(k) = \omega_a - 2J\cos(k)$, and the case of dimerized waveguides, if $\delta\neq 0$, where the model displays two energy bands (see Fig~\ref{fig:schematics}(b)) and where the waveguide may have a non-trivial topology if $\delta<0$ (according to our choice of the unit cell).

The emitters are coupled to the waveguide through an electric-dipole transition appearing between a ground and an optically excited state, which has a relative detuning $\Delta$ with respect to the characteristic bath energy $\omega_a$. In the simplest scenario, these two levels define a spin, whose dipole operator $\s{-}{n}$ exchanges excitations with the waveguide through the usual rotating-wave Hamiltonian $ \sim \sum_n (a^\dagger \s{-}{n}+\mathrm{H.c.})$; $a$ ($a^\dagger$) creates (annihilates) a photon in the waveguide~\cite{CohenTannoudji1998}. Assuming that the emitter's transition energy lies in the bandgap, and that the bath timescales are much faster than the emitter ones (Born-Markov conditions)~\cite{CohenTannoudji1998}, one can adiabatically eliminate the waveguide photons and obtain a coherent spin exchange interaction of the form $\sim \s{+}{n}\s{-}{m}+\mathrm{H.c.}$~\cite{douglas15a,Gonzalez-Tudela2015b} for all emitter pairs. However, in atomic systems one can also define these spins with two hyperfine levels of the ground state manifold, and use Raman-assisted processes to couple them to the waveguide. In this way, one can obtain more general light-matter interaction Hamiltonians, e.g., $\sim a^\dagger \s{x}{n}+\mathrm{H.c.}, a^\dagger \s{z}{n}+\mathrm{H.c.}$, etc., depending on the Raman configuration chosen~\cite{douglas15a,Gonzalez-Tudela2015b,Hung2016}, that will result in different spin interaction processes ($\s{x}{n} \s{x}{m}$, $\s{z}{n}\s{z}{m}$, \dots). Thus, for the sake of generality and to compare with previous works about other systems~\cite{hauke10c}, we will study the emergent many-body phases of the following generalized XXZ model:
\begin{widetext}
  \begin{equation}
    H=-\sum_{j}\sum_{n\geq 0}\sum_{\alpha,\,\beta}
    J^{\alpha\beta}_{n}\left[\frac{\sin\theta}{2}
    \left(\s{+}{j,\alpha}\s{-}{j+n,\beta}+\mathrm{H.c.}\right)
    +\cos\theta\s{z}{j,\alpha}\s{z}{j+n,\beta}\right]
    -\mu\sum_{\alpha}\sum_{j}\s{z}{j,\alpha}\,, \label{eq:Hamiltonian}
  \end{equation}
\end{widetext}
written here in terms of spin-1/2 operators $\s{\nu}{j,\alpha}$, $\nu\in\{x,y,z\}$, and ladder operators $\s{\pm}{j,\alpha}=\s{x}{j,\alpha}\pm i\s{y}{j,\alpha}$. Latin subindices label different unit cells while Greek subindices label different sublattices $\alpha,\beta\in\{A,B\}$ 
[see Fig.~\ref{fig:schematics}(a) for a schematic]. For convenience, we will use sometimes a single subindex to label the spins in the chain; the correspondence between both labellings is given by $\s{\nu}{2j}\equiv\s{\nu}{j,A}$ and $\s{\nu}{2j+1}\equiv\s{\nu}{j,B}$. This generalized model includes:
\begin{itemize}
    \item A tunable weight between the spin exchange and ZZ interactions parametrized by an angle $\theta$, which can be physically obtained by tuning the weight of different Raman transitions~\cite{douglas15a,Gonzalez-Tudela2015b,Hung2016}. For example, $\theta=\pi/2$ corresponds to standard exchange interactions that appear naturally in photon mediated interactions, while $\theta=\pi/4$ corresponds to the isotropic Heisenberg model.
    
    \item A parameter $\mu$ which plays the role of an external magnetic field or chemical potential that fixes the number of excitations in the ground state. Physically, this would correspond either to the energy of the optical transition, if one defines the spin with the emitter's ground and excited states; or to a tunable energy (through the Raman lasers' frequencies) if one defines the spin using the emitter's hyperfine levels.
    
    \item Waveguide-mediated interactions with coupling constants $J_{n}^{\alpha\beta}$, which determine how spins $n$ unit cells apart interact. For simplicity, we have assumed that both the exchange and ZZ terms have the same functional form. Importantly, the functional form of $J_n^{\alpha\beta}$ depends on the particular bandgap the emitters are in resonance with. For a nonzero dimerization ($\delta\neq 0$) the energy spectrum of the SSH photonic bath displays three bandgaps in the single-excitation sector spanning the ranges: $(-\infty,-2J)$, $(-2J\abs{\delta},2J\abs{\delta})$ and $(2J,\infty)$, which we label as the lower (LBG), middle (MBG) and upper (UBG) bandgap, respectively, as depicted in Fig.~\ref{fig:schematics}(b). Each one gives rise to qualitatively 
    different coupling constants $J_{n}^{\alpha\beta}$ as shown in Fig.~\ref{fig:schematics}(c). There, we observe how it is possible to go from long-range antiferromagnetic interactions in the LBG, to fully dimerized couplings in the MBG, and staggered-sign, long-range hoppings in the UBG, which demonstrates the potential of this platform to simulate a wide variety of spin models. 
\end{itemize}

The exact form of $J_{n}^{\alpha\beta}$ was calculated in Ref.~\cite{Bello2019}. Interestingly, they can be parametrized by just two parameters (see Appendix~\ref{sec:couplings}): the interaction length $\xi\in[0,\infty)$, and the effective dimerization constant 
\begin{equation}
    \deltaeff \equiv \frac{\abs{J^{AB}_{n}} - \abs{J^{BA}_{n + 1}}}
  {\abs{J^{AB}_{n}} + \abs{J^{BA}_{n + 1}}} \in [-1, 1] \,.
\end{equation}
For convenience, we also define the ratio 
between coupling constants connecting spins in the same/different sublattice,
\begin{equation}
    \eta \equiv\frac{2\abs{J^{AA}_{n+1}}}{\abs{J^{AB}_n} 
    + \abs{J^{BA}_{n+1}}} = [e^{-1/\xi}(1 - \deltaeff^2)]^{1/2} \in [0, 1]
\end{equation}
. In terms of these parameters, the coupling constants read:
\begin{gather}
  J^{AB}_n = -\Jeff(1 + \deltaeff)e^{-n/\xi}\,, \label{eq:JAB}\\
  J^{BA}_n = -\Jeff(1 - \deltaeff)e^{-(n-1)/\xi}\,,\\
  J^{AA/BB}_n = \Jeff\sign(\Delta)\eta e^{-(n-1)/\xi} \,,
\end{gather}
for $\abs{\Delta}>2J$ (outer bandgaps, UBG and LBG), while
\begin{gather}
  J^{AB}_n = \Jeff\sign(\deltaeff) (1+\deltaeff)(-1)^ne^{-n/\xi}\,,\\
  J^{BA}_n = -\Jeff\sign(\deltaeff)(1-\deltaeff)(-1)^{n-1}e^{-(n-1)/\xi}\,,\\
  J^{AA/BB}_n = \Jeff\sign(\Delta)\eta(-1)^{n-1} e^{-(n-1)/\xi} \,,
  \label{eq:JAA}
\end{gather}
for $\abs{\Delta}<2J\abs{\delta}$ (MBG). We define $J^{\alpha\beta}_0=0$ for $\alpha\beta\in\{AA,BB,BA\}$. The values of $\xi$ and $\deltaeff$ can be tuned to a great extent by tuning different system parameters, such as the emitter's detuning $\Delta$ or the waveguide dimerization parameter $\delta$. However each value of the pair $(\xi,\deltaeff)$ is possible only in either the outer bandgaps or in the inner bandgap, but not both. In the outer bandgaps $\abs{\deltaeff} > (1 - e^{-1/\xi})/(1 + e^{-1/\xi})$, whereas in the inner bandgap the same inequality is fulfilled, replacing ``$>$'' by ``$<$''. Consequently, in the outer bandgaps, interactions connecting spins in the same sublattice are weaker than those connecting spins in different sublattices, $\eta<(1-\abs{\deltaeff})$; however, in the MBG, their strength lies in between, $(1-\abs{\deltaeff})<\eta<(1+\abs{\deltaeff})$, so the system resembles a  ladder rather than a simple 1D chain.

In the next sections, we will analyze the many-body phases emerging from this generalized spin Hamiltonian,
starting in Section~\ref{sec:standard} with the case of a standard waveguide, where the dimerization parameter $\delta=0$, and there only exist the LBG and UBG. Then, in Section~\ref{sec:topological} we move to the more complex scenario with $\delta\neq 0$. But before doing so, we point out to the reader two important symmetries of the model: The Hamiltonian is invariant under rotations about the z-axis of all the spins forming the chain, consequently, the total magnetization $m=\sum_{j,\alpha}\s{z}{j,\alpha}$ is conserved, and we can reduce the problem of finding the ground state of the system to the calculation of the ground state within each fixed-magnetization subspace. Furthermore, it suffices to do so for $\mu=0$, since the chemical potential term is proportional to the identity operator in each fixed-magnetization subspace (the eigenstates of $H$ are the same for all values of $\mu$ and the eigenenergies depend linearly on $\mu$ at an $m$-dependent rate). Another symmetry of the model is the invariance of the Hamiltonian under the simultaneous flip of all spins accompanied by the change $\mu \to - \mu$. Thus, we can limit the study to the phases that appear for $\mu>0$.

\section{Many-body phases in standard waveguide QED \label{sec:standard}}

In this section we analyze the ground states of the generalized XXZ Hamiltonian in the case where the waveguide mediating the interactions has no dimerization ($\delta=0$). In that case, the MBG disappears, and the waveguide-mediated interactions in the LBG and UBG simplify to:
\begin{gather}
  J^{AB}_n = -\Jeff e^{-n/\xi}\,,\\
  J^{BA}_n = -\Jeff e^{-(n-1)/\xi}\,,\\
  J^{AA/BB}_n = \Jeff\sign(\Delta) e^{-1/(2\xi)} e^{-(n-1)/\xi} \,.
\end{gather}
Alternatively, if we use a single-index labelling of the spins along the chain, the coupling constant giving the strength of the interaction between the $i$th and $j$th spin is
\begin{equation}
  J_{ij} = -\Jeff[-\sign(\Delta)]^{\abs{i-j}-1}e^{\frac{1-\abs{i-j}}{2\xi}}\,, \ 
  i\neq j\,.
\end{equation}
As expected, in this case the couplings are not chiral ($\deltaeff= 0$). In this regime, the main differences of the waveguide QED simulator as compared with other analog spin simulators, are the possibility of obtaining longer-range interactions, since $\xi$ can be tuned to be very large \cite{douglas15a}, and the appearance of staggered-sign interactions in the UBG. In the following we first analyze what happens in the limiting cases of nearest-neighbor (NN) (in \ref{subsect:standardnn}) and infinite-range interactions (in \ref{subsect:standardinf}), which are exactly solvable, and will help us better understand the phases that appear for intermediate-range interactions, studied in \ref{subsect:standardint}. In addition, the case of infinite-range interactions is interesting because it connects with the spin models that can be obtained in cavity QED setups~\cite{ritsch13a,davis2019}. 

\subsection{Nearest-neighbour models~\label{subsect:standardnn}}

\begin{figure}
  \includegraphics{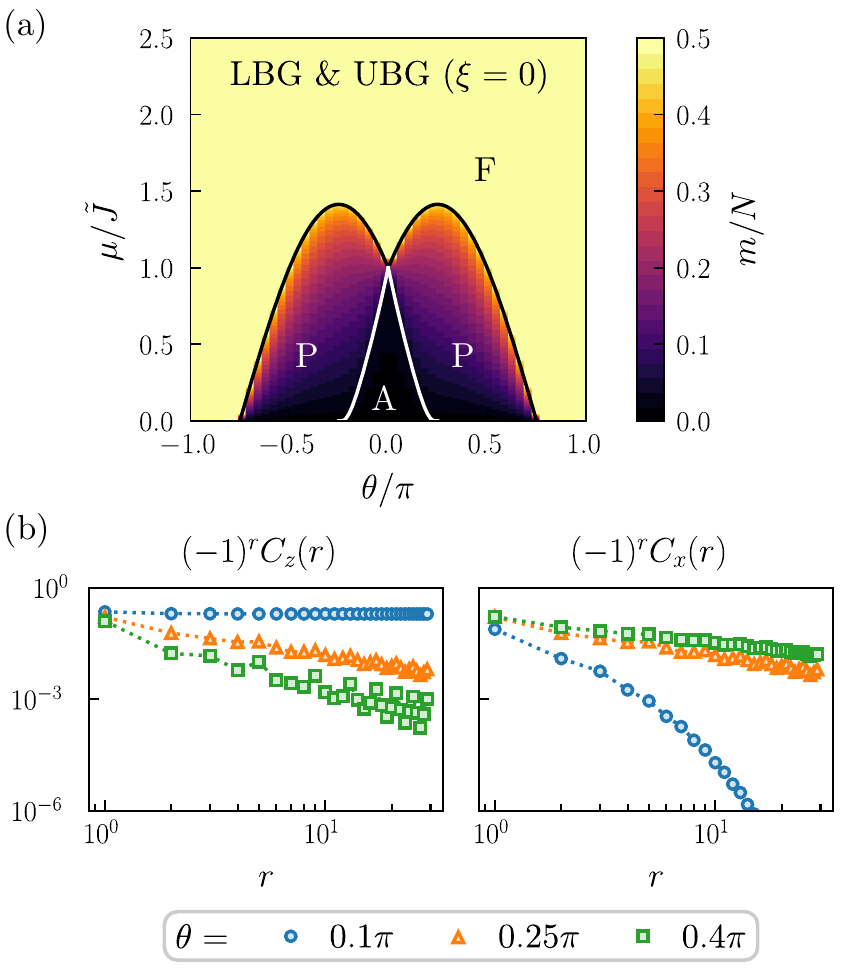}
  \caption{\textbf{Nearest-neighbor XXZ model}. (a) Ground state magnetization, $m\equiv\mean{\sum_{j,\alpha}\s{z}{j,\alpha}}$, as a function of the external magnetic field and the anisotropy angle for the nearest-neighbor XXZ spin model. The results have been obtained using DMRG for a chain with $N=50$ spins. On top of it we plot the analytically-obtained boundaries (black and white lines) of the antiferromagnetic (A), ferromagnetic (F), and planar (P) phases in the thermodynamic limit. (b) Quantum correlations $C_\nu(r)=\mean{\s{\nu}{N/2}\s{\nu}{N/2+r}} - \mean{\s{\nu}{N/2}}\mean{\s{\nu}{N/2+r}}$ in the ground state with zero magnetization ($m=0$) and different values of the anisotropy angle. In the antiferromagnetic phase ($\theta=0.1\pi$) the correlations in the $z$-axis reach a constant value, while the correlations in the $xy$-plane decay exponentially. At the isotropic point ($\theta=\pi/4$), the correlations in any spin direction are the same, and they decay algebraically. In the planar phase ($\theta=0.4\pi$) correlations also decay algebraically, but they are larger in the xy-plane than in the z-axis.\label{fig:NNXXZ}}
\end{figure}

The limit of NN interactions is the same for both the models in the upper and lower bandgaps: $J^{AB}_0=J^{BA}_1=-\Jeff$, while the rest of the couplings are zero. It occurs when $\Delta$ is far-detuned from the bath's energy band. This limit corresponds to the well-known XXZ Heisenberg model, an integrable model which can be solved by Bethe Ansatz techniques \cite{Shollwock2004}. Its phase diagram is shown in Fig.~\ref{fig:NNXXZ}(a). Three distinct phases can be distinguished: 
 
(1) \emph{Ferromagnetic phase}: the ground state is the state in which all spins are aligned in the $z>0$ direction for $\mu > 0$, which we can pictorially represent as $\up\up\up\up\dots$. It is a product state and, therefore, it features no quantum correlations whatsoever between any two spins. The elemental excitations correspond to a single spin flip, $\up\up\down\up\up\dots$. If the xx- and yy-interaction terms are nonzero, these excitations can hop between neighbouring spins, so the eigenstates of the $m=N/2-1$ subspace are spin waves that carry spin 1, also known as \emph{magnons}. Assuming $\mu>0$, the single magnon spectrum is given by
\begin{equation}
  \frac{\epsilon(k)}{\Jeff}=\sin\theta\,\cos k+\cos\theta\left(\frac{N}{4}-1\right)
  -\frac{\mu}{\Jeff}\left(\frac{N}{2}-1\right)\,,
\end{equation}
where $k$ is the magnon quasimomentum. Computing the minimum of this dispersion relation, one can readily find the energy gap between the ground state and the first excited state as $\Delta E/\Jeff=-\abs{\sin\theta} - \cos\theta + \mu/\Jeff$. The critical values of the model parameters at which there is a phase transition from the ferromagnetic phase to other phases of the model are the points where $\Delta E = 0$ [shown as a black curve in Fig.~\ref{fig:NNXXZ}(a)].

(2) \emph{Néel phase} (aka \emph{antiferromagnetic phase}): At the point $\theta = 0$, for positive zz-interactions, the classical ground state is the state in which spins are pointing in opposite directions in an alternating fashion, $\up\down\up\down\dots$, also known as antiferromagnetic order. The lowest-energy excitations in this phase correspond to \emph{spinons}, fermionic quasiparticles that carry spin 1/2. They correspond to domain walls between the two possible antiferromagnetic orderings $\up\down\up\down\down\up\down\up\dots$. The magnetization can only change in integer steps, so these excitations are created in pairs as the magnetization increases. In the vicinity of the Ising limit, for small, finite xx- and yy-interactions, this picture is still valid. The ground state features long-range antiferromagnetic correlations, and the elemental excitations are spinons, which can now hop along the chain by two sites. Using perturbation theory \cite{Ishimura1980}, one can compute the energy of two spinons (states that belong to the subspace with $m=1$) as $\epsilon(k_1)+\epsilon(k_2)$, where 
\begin{equation}
    \frac{\epsilon(k)}{\Jeff}=-\frac{\cos\theta}{4}(N-2) + \sin\theta \cos 2k - \frac{\mu}{2\Jeff}\,.
\end{equation}
They form a continuum of scattering eigenstates. Therefore, close to the classical antiferromagnetic Ising model ($\theta\to 0$) the energy gap is $\Delta E/\Jeff\approx \cos\theta - 2\abs{\sin\theta} - \mu/\Jeff$, which gives a good approximation of the critical line shown in Fig.~\ref{fig:NNXXZ}(a) in white. The exact energy gap in the thermodynamic limit can be obtained using the Bethe Ansatz \cite{DesCloizeaux1966}, which is the formula we have used in Fig.~\ref{fig:NNXXZ}(a).

(3) \emph{Planar phase} (aka \emph{XY phase}, or \emph{Tomonaga-Luttinger phase}) This phase is fundamentally different from the previous two, as the system is gapless, i.e., excitations over the ground state can be produced at arbitrary small energies. In these situations the system is said to be in a critical phase. This can be better understood at the XX limit ($\theta\to \pi/2$), where the spin chain can be mapped to a system of free fermions hopping on a 1D lattice via the Jordan-Wigner transformation \cite{Lieb1961} (Appendix \ref{sec:dimerXX}). These fermions have the familiar dispersion relation $\epsilon(k)=2\Jeff\cos(k)-\mu$, and the ground state can be constructed as the Slater determinant of all single-particle states with energies $\epsilon(k)<0$. For $\abs{\mu}<2\Jeff$, the Fermi level lies somewhere in the middle of the energy band and excitations can be created adding fermions with energies above the Fermi level (possibly removing them from energy states below it). Being a critical phase, correlations decay as a power law (algebraically), whereas in gapped phases they decay exponentially with increasing distance, see Fig.~\ref{fig:NNXXZ}(b). Correlations in this phase occur predominantly in the xy-plane. Away from the XX limit the Jordan-Wigner transformation maps the spin chain to a model or interacting fermions in 1D, which is the subject of the so-called Tomonaga-Luttinger theory \cite{Giamarchi2003quantph1d}.

Since the system is bipartite---interactions only involve spins that belong to different sublattices---the phase diagram is symmetric with respect to the change $\theta\to-\theta$. To understand why, note that a transformation with $U=\prod_{j}i2\s{z}{j,A}$ inverts the sign of the flip-flop terms of the Hamiltonian, while it preserves the total magnetization. 

\subsection{Infinite-ranged interactions~\label{subsect:standardinf}}

\begin{figure}[tb]
    \centering
    \includegraphics{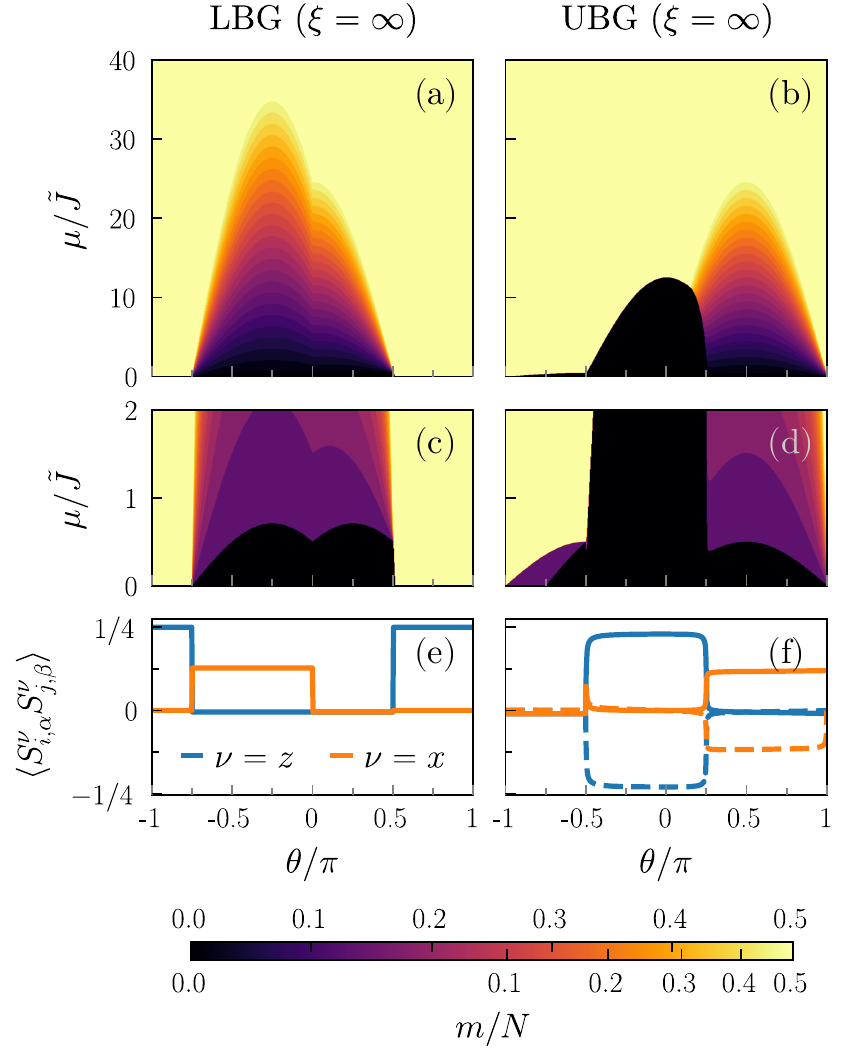}
    \caption{\textbf{Undimerized, infinite interaction-range models in the LBG and UBG}. (a, b) Ground-state magnetization phase diagrams for a chain with $N=50$ spins. (c, d) Zoom into the phase diagrams for small values of $\mu$. In order to distinguish the different magnetization sectors, the color scale used (ticks under the color bar) is different from the one used in the upper panels (ticks above the color bar). (e, f) Correlations in the ground state for $\mu\to 0^+$; the continuous (dashed) lines in panel (f) show the correlations between spins that belong to the same (opposite) sublattice.}
    \label{fig:infinite_range}
\end{figure}

Adding couplings beyond nearest neighbours breaks the sublattice symmetry (the system is no longer bipartite), making the phase diagram asymmetric with respect to the change $\theta\to-\theta$. This can be clearly appreciated in the limit of infinite-range interactions, Fig.~\ref{fig:infinite_range}(a, b). In this limit, the dimensionality of the model is irrelevant, and the ground state can be found using collective spin operators. These infinite-range models occur in the upper and lower bandgaps as the emitter transition frequency gets closer to the band edges ($\Delta\to-2\Jeff$ for the LBG, and $\Delta\to 2\Jeff$ for the UBG). We should remark however that the Markovian approximation becomes worse as the frequency of the emitters gets closer to the band edges, so this limit would be hard to achieve in actual experiments. In any case, understanding the physics of these models will allow us to better comprehend what happens at intermediate interaction ranges.

In the LBG, the coupling constants of the infinite-range model are
$J^{\alpha\beta}_n=-\tilde J$, and the Hamiltonian can be written, apart 
from a constant term, as
\begin{equation}
  H = \frac{\Jeff}{2} \left[ 
  \frac{\sin\theta}{2} \left(S^+ S^- + S^- S^+\right)
  +\cos\theta \left(S^z\right)^2 \right] - \mu S^z\,,
\end{equation}
with $S^\nu \equiv \sum_{n,\alpha} \s{\nu}{n,\alpha}$.
This Hamiltonian is diagonal in the basis of Dicke states $\ket{s,m,\lambda}$, which
are eigenstates of $\mathbf S^2$ and $S^z$ with eigenvalues $s(s+1)$ and $m$
respectively \cite{arecchi72a}. The energy of these states is given by
\begin{equation}
  E(s,m) = \frac{\tilde J}{2}\left\{\sin\theta\left[s(s + 1) - m^2\right]
  + \cos\theta\, m^2\right\} - \mu m\,.
\end{equation}
The phase diagram shown in Fig.~\ref{fig:infinite_range}(a) can be obtained 
minimizing $E(s,m)$. For any fixed magnetization $m$, the ground state has 
maximum total angular momentum $s=N/2$ for $\theta<0$, while it has minimum 
total angular momentum compatible with the given magnetization $s=m$ in the 
case $\theta>0$. In the classical limits ($\theta=0,\pi$), the Hamiltonian is
diagonal in the basis of Ising configurations so the energy becomes 
independent of $s$. Correlations between different spins in the case 
$\theta<0$ can be easily computed noting that the states with maximum total 
angular momentum are non-degenerate and they are invariant under any 
permutation of the spins in the chain. Thus, the correlations are the same 
regardless the distance separating the spins
\begin{align}
  \mean{\s{z}{i,\alpha}\s{z}{j,\beta}} & =\frac{4m^2-N}{4N(N-1)}\,, \label{eq:corrzz}\\
  \mean{\s{x}{i,\alpha}\s{x}{j,\beta}} & =
  \mean{\s{y}{i,\alpha}\s{y}{j,\beta}} = \frac{N^2-4m^2}{8N(N-1)}\,,
\end{align}
for $(i,\alpha)\neq(j,\beta)$. For $\theta>0$, the correlations are not uniquely determined, since the ground state is degenerate. This degeneracy is associated with the permutation quantum number $\lambda$ \cite{arecchi72a}. Nevertheless, if we consider a fully mixed state within the ground-state manifold (a thermal state $\rho\propto e^{-\beta H}$ in the limit $\beta\to\infty$) the correlations become invariant under any permutation of the spins forming the chain; the zz-correlations in this case are the same as the ones for $\theta < 0$ [Eq.~\eqref{eq:corrzz}], but the xx- and yy-correlations are different, they are given by
\begin{align}
  \mean{\s{x}{i,\alpha}\s{x}{j,\beta}} & =
  \mean{\s{y}{i,\alpha}\s{y}{j,\beta}} = \frac{2m-N}{4N(N-1)}\,.
\end{align}
In the infinite interaction-range models the energy needed to flip a spin becomes an extensive quantity, since each spin interacts equally strongly with all the other spins forming the chain. Consequently, the stability region of each phase with a fixed magnetization $m$ shifts to higher values of the external magnetic field $\mu$ as we increase the number of spins forming the chain. Eventually, as we approach the thermodynamic limit, the ground state phases for any finite $\mu$ correspond to those appearing in the limit $\mu\to 0^+$, see Fig.~\ref{fig:infinite_range}(c). The correlations in these phases are shown for a chain with $N=50$ spins in Fig.~\ref{fig:infinite_range}(e).

The coupling constants of the infinite-range model in the UBG are $J^{AB/BA}_n=-J^{AA/BB}_n=-\tilde J$. The Hamiltonian can be written in this case as 
\begin{multline}
    H = -\frac{\Jeff}{2} \sum_{\alpha=A,B} \left[ 
      \frac{\sin\theta}{2} \left(\s{+}{\alpha}\s{-}{\alpha} 
      + \s{-}{\alpha} \s{+}{\alpha}\right) 
      + \cos\theta \left(\s{z}{\alpha}\right)^2 \right] \\
      + \Jeff \left[ \frac{\sin\theta}{2} \left(\s{+}{A}\s{-}{B}
      + \s{-}{A}\s{+}{B}\right) + \cos\theta \s{z}{A}\s{z}{B}\right] \\
      - \mu\left(\s{z}{A} + \s{z}{B}\right) \,,
    \label{eq:infinite_range_upper}
\end{multline}
with $\s{\nu}{\alpha}\equiv\sum_n\s{\nu}{n,\alpha}$. This Hamiltonian preserves the total angular momentum in each sublattice $s_A$ and $s_B$, and the total magnetization $m$. Its matrix representation in the subspace with fixed $(s_A, s_B, m)$ can be cast into tridiagonal form (see appendix~\ref{sec:IRupper}), so it can be easily diagonalized for a large number of spins. In Fig.~\ref{fig:infinite_range}(b) we show the phase diagram obtained for a chain with $N=50$ spins. We find that for $-\pi/2<\theta<\pi$ the ground state is unique and is fully symmetric under permutations of spins within each sublattice. Thus, correlations do not depend on the distance separating the spins, but are different for spins within the same/opposite sublattice, see Fig.~\ref{fig:infinite_range}(f). In some cases, the ground state can be computed analytically. For example, at the isotropic (Heisenberg) point $\theta=\pi/4$ the ground state for $\mu=0$ is a singlet of total angular momentum $\mathbf S^2=(\mathbf S_A + \mathbf S_B)^2$, and the correlations are 
\begin{equation}
  \mean{\s{\nu}{i,A}\s{\nu}{j,B}} = -\frac{N + 4}{12 N}
  \,,\ \mean{\s{\nu}{i,\alpha}\s{\nu}{j,\alpha}} = \frac{1}{12} \,.
\end{equation}

For $-\pi<\theta<\pi$ the correlations are not uniquely determined, since the ground state is degenerate. Again, if we consider a thermal state in the zero-temperature limit, correlations become homogeneous and can be computed analytically (see appendix \ref{sec:IRupper}).

\subsection{Waveguide-mediated interactions: intermediate ranges~\label{subsect:standardint}}

\begin{figure*}[tb]
    \centering
    \includegraphics{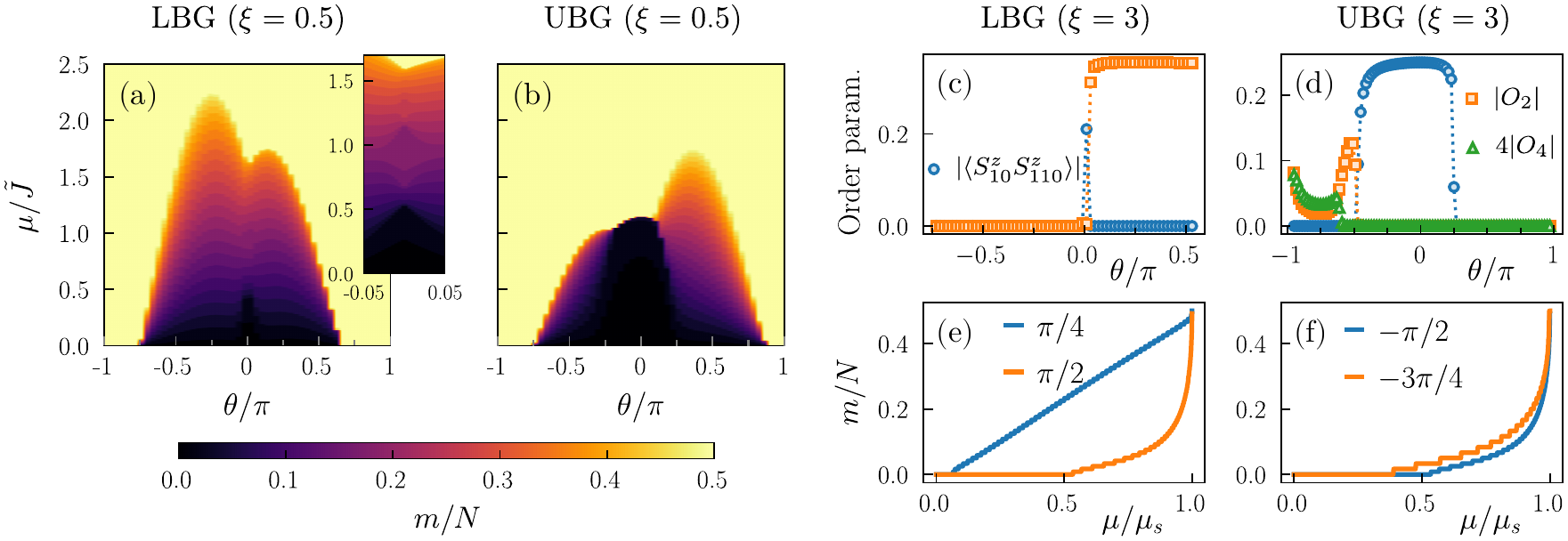}
    \caption{\textbf{Undimerized, intermediate interaction-range models in the LBG and UBG} (a, b) Ground-state magnetization phase diagrams for a chain with $N=50$ spins. (c, d) Different order parameters as a function of the anisotropy angle in the ground state with zero magnetization, for a chain with $N=120$ spins (the legend is the same for both plots, in the LBG the values of $|O_4|$ computed are negligible). (e, f) Magnetization curves for a system with the same parameters in the pannels (c, d), for particular values of $\theta$, as indicated in the legend. Here, $\mu_s$ denotes the saturation field, i.e., the value of $\mu$ at which the system becomes fully polarized. Note that this $\mu_s$ is a function of $\theta$. \label{fig:undimerized_intermediate}}
\end{figure*}

For intermediate interaction lengths we may expect the phase diagram to interpolate between that of the nearest-neighbour XXZ Heisenberg model, and the ones in the infinite-range interaction limit. This can be clearly appreciated in Fig.~\ref{fig:undimerized_intermediate}(a, b) [cf.~Fig.~\ref{fig:NNXXZ} and Fig.~\ref{fig:infinite_range}(a, b)]. In the light of these results, we can see that increasing the interaction length in the LBG shrinks the antiferromagnetic phase, confining it to values of $\theta$ closer to the classical limit ($\theta=0$), while in the UBG it enlarges it, making this phase more stable against quantum fluctuations. This is confirmed also by the long-range zz-correlations, see Fig.~\ref{fig:undimerized_intermediate}(c, d).

However, this is not the end of the story, as there are other phases that appear only when the interaction range is large but finite. For example, in the LBG, the phase diagram develops a series of magnetization plateaus [see inset of Fig.~\ref{fig:undimerized_intermediate}(a)] around the classical regime $\theta=0$, which in the thermodynamic limit have a fractal structure known as \emph{Devil's staircase} \cite{hauke10c,bak82a}. These plateaus correspond to Ising-like phases with periodic zz-correlations with  different periods (patterns). 

Also, in the LBG, for positive exchange interactions ($0<\theta<\pi$) there is a phase transition in the sector with zero total magnetization (small values of $\mu$) from a planar phase to a gapped, spontaneously dimerized phase. This effect of frustration has been extensively studied in the past in simpler models, such as the antiferromagnetic zigzag Heisenberg chain \cite{haldane1982,hikihara2001,okunishi2003,mcculloch2008,sandvik2010}. A model where this phenomenon can be easily understood is the exactly-solvable Majumdar-Ghosh model \cite{auerbachbook94a}, whose ground state is a product of singlets between adjacent spins. Phases of this kind are known as valence-bond solid (VBS) phases. Generally, these phases are gapped phases that do not feature any long-range order, i.e., the correlations $\mean{S^\nu_i S^\nu_j}$ decay exponentially with increasing distance $\abs{i - j}$. However, fixing the distance, they become periodic as different spins along the chain are considered. In the spontaneously-dimerized VBS phase appearing in the LBG, the correlations between neighbouring spins (bond correlations) have period $p=2$, $\mean{B^\nu_j}\equiv\mean{S^\nu_jS^\nu_{j+1}} \simeq \overline{B^\nu} + (-1)^j\delta B^\nu$ \footnote{For closed boundary conditions the two possible dimerization 
patterns are degenerate, and the observed $\delta B^\nu$ may be zero. 
Still, this phase is characterized by the long-range order of the 
\emph{bond-bond correlation functions} $D(r)=\mean{B^\nu_nB^\nu_{n+r}}$}. 
This dimer order in the bond correlations, as well as other orders with a larger spatial periodicity, can be detected experimentally by measuring spin correlations along the chain and computing the following quantity (bond-order parameter or structure factor): 
\begin{equation}
  O_p=\frac{1}{L}\sum_{n=n_0}^{n_0 + L - 1}
  \mean{\vs{n}\vs{n+1}}e^{-i\frac{2\pi}{p}n}\,.
  \label{eq:bondorder}
\end{equation}
In particular, the spontaneous dimerization can be quantified by $\abs{O_2}$. In Fig.~\ref{fig:undimerized_intermediate}(c, d) we plot this order parameter showing that dimer VBS phases occur at magnetization $m=0$ for $0<\theta<\pi/2$ in the LBG, and $-\pi<\theta<-\pi/2$ in the UBG. For higher magnetization, $m\neq 0$, the value of $\abs{O_2}$ is negligible. 

Curiously, in the UBG we also find a VBS phase with a doubled period (tetramer order) for values of the anisotropy angle $\theta\approx -3\pi/4$,
see Fig.~\ref{fig:undimerized_intermediate}(d). Furthermore, in the magnetization curve shown in Fig.~\ref{fig:undimerized_intermediate}(f), we observe that the magnetization changes in steps of $\Delta m=2$ as $\mu$ increases. This is in contrast with the other cases studied, in which the magnetization increases in steps of $\Delta m=1$, cf.~\ref{fig:undimerized_intermediate}(e). Considering that in this case the couplings between spins in opposite sublattices are ferromagnetic, this suggest that excitations correspond in this tetramer phase to simultaneous flips of two spins, one in each sublattice. These gapless phases correspond to spin-density wave (SDW) and nematic phases for small and large magnetization respectively, already identified in the zigzag Heisenberg chain with ferromagnetic NN interactions and antiferromagnetic next-nearest-neighbor (NNN) interactions \cite{hikihara2008} (see appendix~\ref{sec:nematic}).


\section{Many-body phases in topological waveguide QED \label{sec:topological}}

\subsection{Dimerized 1D photonic bath (upper and lower bandgaps)}

\begin{figure}[tb]
  \includegraphics{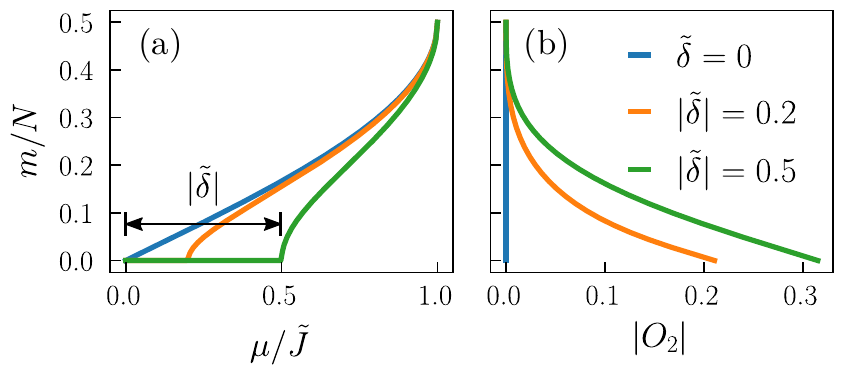}
  \caption{\textbf{Dimerized NN-XX model} ($\theta=\pi/2$). (a) Magnetization curves for different values of the effective dimerization. (b) Dimer VBS order parameter as a function of the magnetization for several values of $\abs{\deltaeff}$ [same as in (a)]. 
  \label{fig:dimerXX}}
\end{figure}

We now analyze the cases where $\deltaeff \neq 0$. Again, we start by considering the limit $\xi\to 0$, i.e., only NN couplings. As we shall see, the dimerization produces VBS phases, even though there is no frustration in the system. This can be better understood in the XX models, $\theta=\pm\pi/2$, which can be mapped exactly to a free fermionic SSH model by a Jordan-Wigner transformation (see appendix~\ref{sec:dimerXX}). A nonzero dimerization of the interactions opens a gap in the energy spectrum between the ground state with zero magnetization ($m=0$) and the ground states of other magnetization sectors ($m\neq 0$), see Fig.~\ref{fig:dimerXX}(a). Correlations become dimerized with a dimerization pattern that follows that of the underlying photonic SSH bath. In contrast with the models discussed in the previous section, the phases with nonzero magnetization also feature a nonzero value of $\abs{O_2}$ that decreases as the magnetization increases, see Fig.~\ref{fig:dimerXX}(b).

\begin{figure}[tb]
  \includegraphics{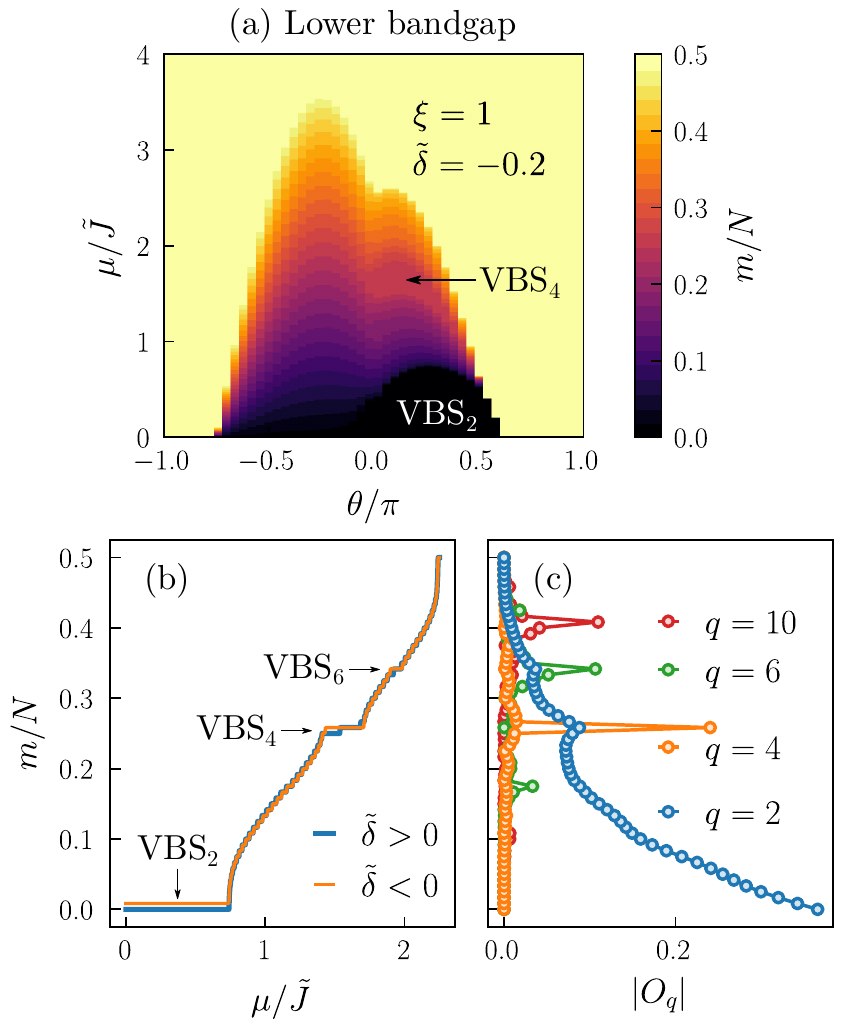}
  \caption{\textbf{Dimerized, intermediate interaction-range, LBG model.} (a) Ground-state magnetization as a function of the external 
  magnetic field and the anisotropy angle for a system with $N=50$ spins.
  (b) Magnetization curve for a longer chain with $N=120$ for 
  $\theta=\pi/4$ and $\deltaeff=\pm 0.2$. 
  (c) Bond order parameters (structure factor) for the model with $\theta=\pi/4$ and 
  $\deltaeff=0.2$ as a function of the magnetization. In all the figures 
  the interaction length is set to $\xi=1$. \label{fig:dimerized_lower}}
\end{figure}

\begin{figure}[tb]
  \centering
  \includegraphics{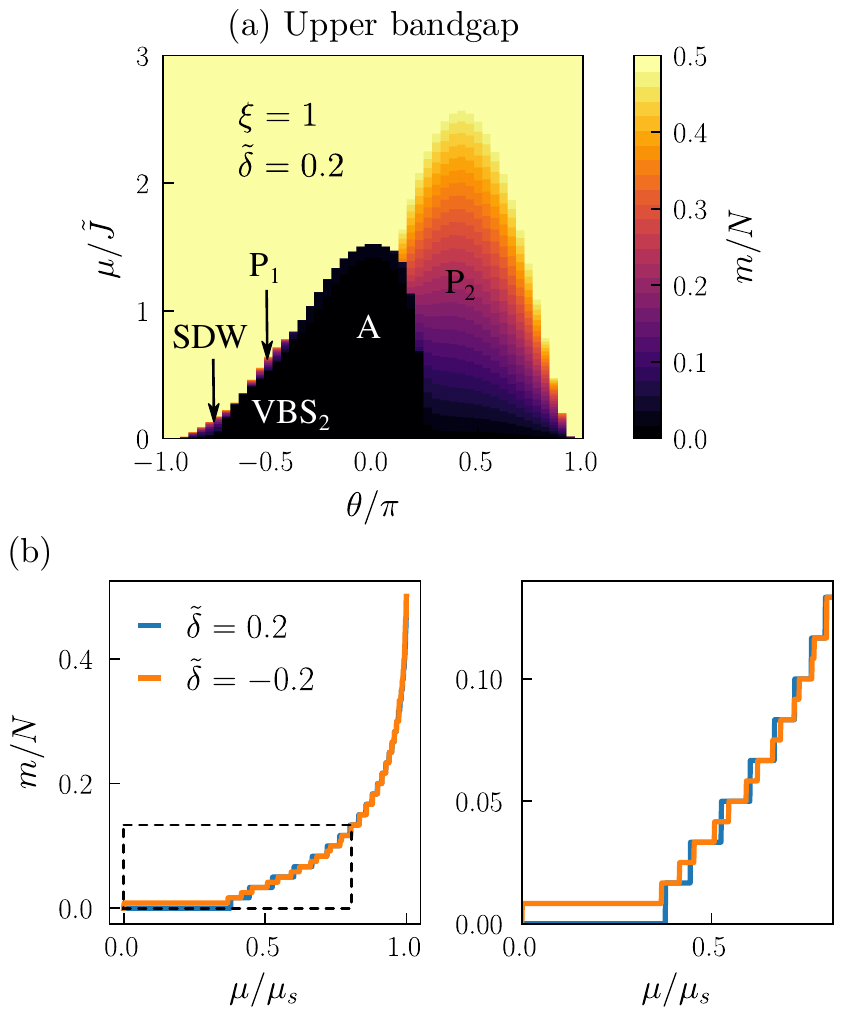}
  \includegraphics{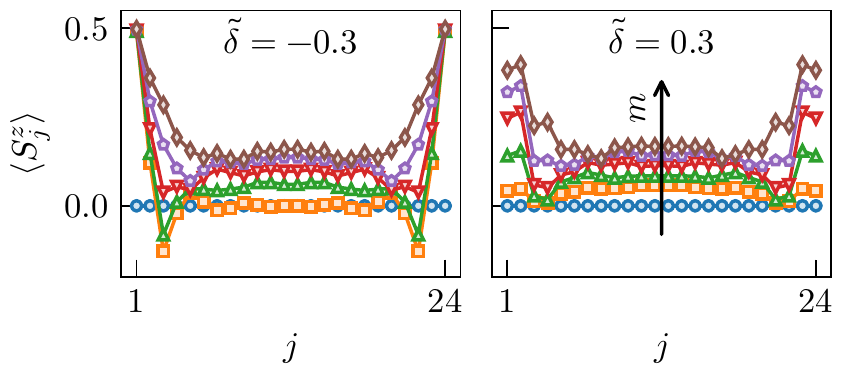}
  \caption{\textbf{Dimerized, intermediate interaction-range, UBG model.} (a) Ground-state magnetization as a function of the external 
  magnetic field and the anisotropy angle for a system with $N=50$ spins.
  (b) Magnetization curve for a longer chain with $N = 120$ for 
  $\theta = -3\pi/4$ and $\deltaeff = \pm 0.2$. The right plot is just a 
  zoom into the lower magnetization range. 
  (c) Individual-spin magnetization for a chain with $N=24$ spins and OBC, at $\theta=-0.65\pi$. The values correspond to the ground states of the sectors with total magnetization $0\leq m\leq 5$. Even though for $\deltaeff > 0$ there is a considerable effect of the edges, in the case with $\deltaeff < 0$ the difference with respect to the bulk is starker, with the ending spins being fully polarized regardless the value of the total magnetization. In all the panels, the 
  interaction length is set to $\xi=1$.\label{fig:dimerized_upper}}
\end{figure}

Since the dimerized NN-XX-model can be mapped to the fermionic SSH model, the topology of the former can be linked straightforwardly to that of the latter. Thus, we expect two distinct symmetry protected topological (SPT) phases at $m=0$: one for $\deltaeff > 0$ (trivial SPT phase) and another one for $\deltaeff < 0$ (non-trivial SPT phase). These SPT phases extend to other values of $\theta$ around the XX point and larger values of $\xi$, see Figs.~\ref{fig:dimerized_lower}(a) and \ref{fig:dimerized_upper}(a) (phase labelled as $\mathrm{VBS}_2$), which in the fermionic language correspond to models of interacting fermions. Since these models can be connected adiabatically to the free-fermion model without closing the gap, they all belong to the same SPT phases. Thus, we can argue that these \emph{non-interacting} SPT phases are robust against some amount of interaction. More specifically, they are robust against correlated fermion hopping (long-range spin interactions) but not so much against fermion density-density interactions (Ising-like spin interactions). 

The distinction between the two possible SPT phases becomes apparent when considering chains with open boundary conditions (OBC). In the non-trivial SPT phase ($\deltaeff < 0$), the ground state is fourfold degenerate in the thermodynamic limit if $\mu = 0$ due to the presence of two uncoupled spin-1/2 degrees of freedom located at the edges of the chain. This degeneracy, and hence the presence of these edge modes, are a consequence of symmetry-fractionalization~\cite{verresen2017}. Both are protected by several symmetries: time-reversal and the group generated by $\pi$-rotations around two orthogonal spin axes. For $\mu\neq 0$, these free spins at the ends of the chain align with the external magnetic field, resulting in a difference of the ground-state magnetization $m(\delta<0)-m(\delta>0)=1$ for $\mu\to 0^+$. This can be clearly appreciated in the magnetization curves shown in Figs.~\ref{fig:dimerized_lower}(b) and \ref{fig:dimerized_upper}(b). 

Formally, these different SPT phases can be distinguished by the geometric phases accumulated by the ground state when modulating periodically (``twisting'') the Hamiltonian in certain specific ways. These many-body Berry phases may be quantized depending on the symmetries of the model and on the specific twist performed~\cite{hatsugai2006,hirano2008,kariyado2018}. For example, for a time-reversal symmetric phase, such as the $\mathrm{VBS}_2$ phase, a modulation of the interaction term between any two spins $m, n \in \{ 1,\dots,N\}$ of the form:
\begin{equation}
    S^+_m S^-_n + \mathrm{H.c.} \to e^{-i\phi}S^+_m S^-_n + \mathrm{H.c.}\,
\end{equation}
as $\phi$ varies through $[0,2\pi)$, will result in a Berry phase equal to 0 or $\pi$ (mod $2\pi$), depending on whether the two spins are weakly or strongly entangled, respectively \cite{hatsugai2006}. The many-body Berry phases computed between neighboring spins thus reveal the pattern of singlets characteristic of each SPT order. We obtain different values depending on whether the two spins belong to the same/adjacent unit cells, $\gamma_\mathrm{intra/inter}$:
\begin{equation}
    (\gamma_\mathrm{intra},\gamma_\mathrm{inter}) = \begin{cases}
    (\pi, 0) & \text{for } \deltaeff > 0\\
    (0, \pi) & \text{for } \deltaeff < 0
    \end{cases} \mod 2\pi\,.
    \label{eq:berry}
\end{equation}

\begin{figure}
    \centering
    \includegraphics{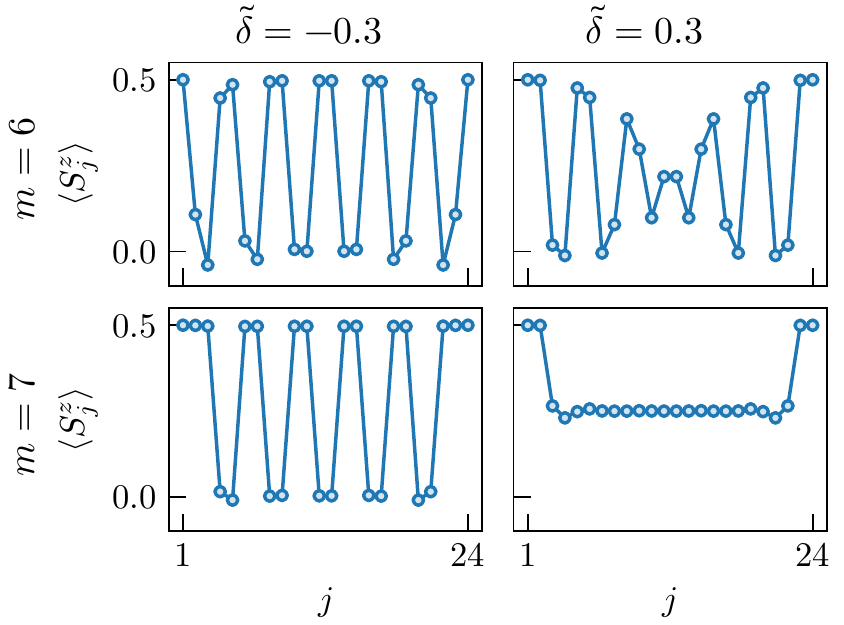}
    \caption{\textbf{Dimerized, intermediate interaction-range, LBG model.} Individual-spin magnetization for chains with $N=24$ spins with OBC at the isotropic point $\theta=\pi/4$. The upper (lower) row corresponds to phases with $m=N/4$ ($m=N/4 + 1$), whereas the left (right) column corresponds to phases with negative (positive) dimerization constant. The interaction length is set to $\xi = 1$ in all cases. In all cases considered, the ground state is unique, except for the case with $\deltaeff > 0$ and $m = N/4 + 1$, where it is doubly degenerate. The bulk of the chain in this case would be the one that resembles most closely the $\mathrm{VBS}_4$ state in chains with PBC}
    \label{fig:vbs4_obc}
\end{figure}

\begin{figure}
    \centering
    \includegraphics[width=\linewidth]{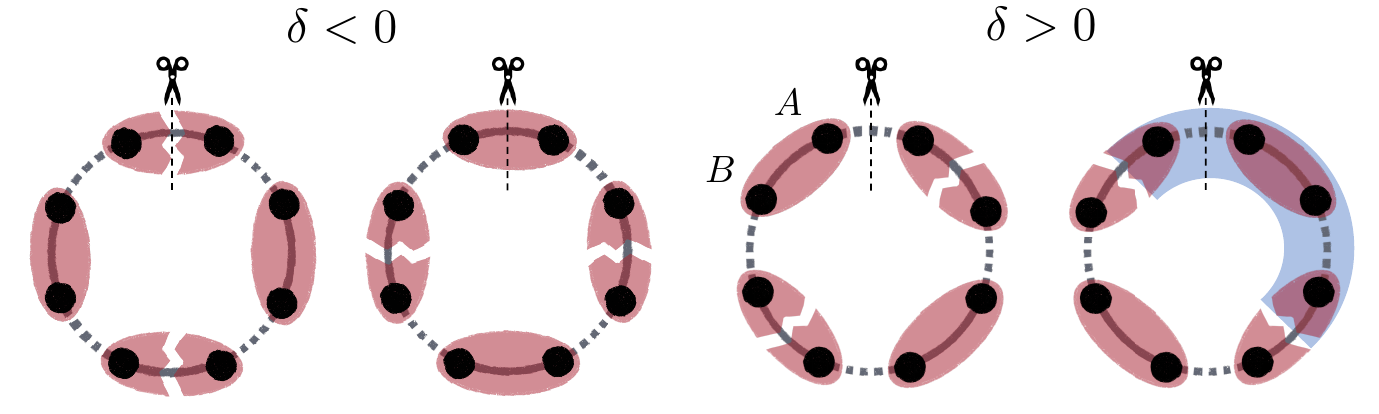}
    \caption{\textbf{Schematic representation of the $\mathrm{VBS}_4$ phase.} Red ellipses represent valence bonds (singlets) between adjacent spins (black dots). Split ellipses represent broken valence bonds in which the two spins are fully polarized along the z-axis. The blue shaded area represents the space-inversion symmetric ``cluster'' that repeats along the chain in this VBS phase. For each distinct SPT phase (corresponding to positive or negative values of the dimerization constant $\delta$), there are two possible patterns of broken bonds, and hence, the ground state of chains with PBC is twofold degenerate. If we consider chains with OBC, we see that for $\delta < 0$ any of the two patterns can be cut open still respecting inversion symmetry (the leftmost pattern is energetically more favorable for $m=N/4$, while the other one is preferred for $m=N/4 + 1$). For $\delta > 0$, by contrast, this is not possible and there is no preferred pattern of broken bonds for chains with OBC, c.f. Fig.~\ref{fig:vbs4_obc}}
    \label{fig:vbs4_schematic}
\end{figure}

Surprisingly, for a finite interaction length, $\xi\neq 0$, the models in the LBG develop new gapped phases at magnetizations $m/N\approx 1/4$ and $m/N\approx 1/3$. These phases correspond to the magnetization plateaus shown in Fig.~\ref{fig:dimerized_lower}(a, b) for $\theta\approx \pi/4$, labelled as $\mathrm{VBS}_q$, for $q = 4, 6$. When we compute the bond-order parameter $O_q$ for different periodicities, we observe that they have a period involving several unit cells, see Fig.~\ref{fig:dimerized_lower}(c). This can also be observed in the individual spin magnetization for chains with OBC, see Fig.~\ref{fig:vbs4_obc}, which suggests the following simplified picture of the ground state: it is formed by ``breaking'' in a periodic fashion some of the valence bonds present in the $\mathrm{VBS}_2$ phase, as shown schematically in Fig.~\ref{fig:vbs4_schematic}. This intuitive picture also explains the ($q/2$)-fold degeneracy of the ground state in the sector with magnetization $m/N=1/2 - 1/q$ found in chains with periodic boundary conditions (PBC), as there are $q/2$ different ways to break the bonds in in order to achieve a phase with that magnetization .

Similarly to the $\mathrm{VBS}_2$ phase, these $\mathrm{VBS}_q$ ($q>2$) phases are also SPT phases that can either be topologically trivial ($\deltaeff > 0$) or not ($\deltaeff < 0$). This is confirmed computing the (non-Abelian) many-body Berry phases (see Appendix~\ref{sec:berry}), which are the same as in the $q=2$ phase, Eq.~\eqref{eq:berry}. However, in contrast with the $q=2$ phase, these SPT phases at higher magnetization are protected solely by space-inversion symmetry. As it turns out, in this case non-trivial chains with OBC do not feature topological edge states \cite{pollmann2010,hughes2011}, and the ground state of trivial chains with OBC presents a twofold degeneracy only in the case with $\deltaeff > 0$ and $m=N/4 + 1$, which is not attributed to the presence of edge modes, but rather to a special kind of frustration related to the lack of inversion symmetry of the two possible patterns of broken bonds, see Fig.~\ref{fig:vbs4_schematic}.

We remark that the appearance of these higher-period VBS phases requires the conjuntion of long-range exchange and zz-interactions, and a non-zero dimerization. In the fermionic language, we can say that they appear due to the presence of interactions between fermions (both of the density-density, and of the correlated-hopping kind).

As for the dimerized UBG model with finite interaction length, a similar analysis follows for the $\mathrm{VBS}_2$ phase located around $\theta\approx -\pi/2$: it is topologically trivial (non-trivial) if $\deltaeff > 0$ ($\deltaeff < 0$). Curiously, we observe that the dimerization affects the SDW phase appearing at $\theta\approx-3\pi/4$ in low external magnetic fields. As shown in Fig.~\ref{fig:dimerized_upper}(b), for $\deltaeff<0$ the magnetization changes in steps of $\Delta m = 1$, instead of $\Delta m = 2$, which is the case for $\deltaeff>0$. This difference suggests the presence of ``persistent edge modes'', that is, almost free spin-1/2 degrees of freedom at the edges of the chain that align in the direction of the external magnetic field, see Fig.~\ref{fig:dimerized_upper}(c).

\subsection{Middle bandgap}                       

For $\delta\neq 0$ a new gap appears in the bath's energy spectrum (MBG). 
In this gap, all spin coupling constants $J^{\alpha\beta}_n$ alternate sign 
as a function of distance $n$. Again, let us consider first the NN and 
infinite-range limits. The NN limit corresponds in this case to a set of 
disconnected dimers. It is obtained for $\Delta=0$ and $\delta\to \pm 1$. 
On the other hand, there are several posible infinite-range interaction models. For 
$\Delta\to \pm 2\abs{\delta}J$ when $\delta\neq 0$, the couplings are 
$J^{AB}_n=\sign(\delta)(-1)^n\Jeff$, 
$J^{BA}_n=-\sign(\delta)(-1)^{n-1}\Jeff$, and 
$J^{AA/BB}_n=\sign(\Delta)(-1)^{n-1}\Jeff$. For $\Delta\to -2\abs{\delta}J$,
the Hamiltonian is the same as the one in 
Eq.~\eqref{eq:infinite_range_upper}
with $\s{\nu}{A}=\sum_{n}\left(\s{\nu}{2n,A}+\s{\nu}{2n,B}\right)$ and 
$\s{\nu}{B}=\sum_{n}\left(\s{\nu}{2n+1,A}+\s{\nu}{2n+1,B}\right)$. Thus, 
the ground state magnetization as a function of the anisotropy angle and the
external magnetic field is exactly the same as the one shown in 
Fig.~\ref{fig:infinite_range}(b). For $\Delta\to 2\abs{\delta}J$, defining 
$\s{\nu}{A}=\sum_{n}\left(\s{\nu}{2n,A}+\s{\nu}{2n+1,B}\right)$ and 
$\s{\nu}{B}=\sum_{n}\left(\s{\nu}{2n+1,A}+\s{\nu}{2n,B}\right)$ the 
Hamiltonian is the same as the one in Eq.~\eqref{eq:infinite_range_upper} 
but with opposite sign. Thus, the 
ground-state magnetization is again the same as the one shown in 
Fig.~\ref{fig:infinite_range}(b), shifting $\theta\to\theta+\pi$. The analysis based on Dicke states is still valid, but note that now the unit cell has doubled.

\begin{figure}[tb]
  \includegraphics{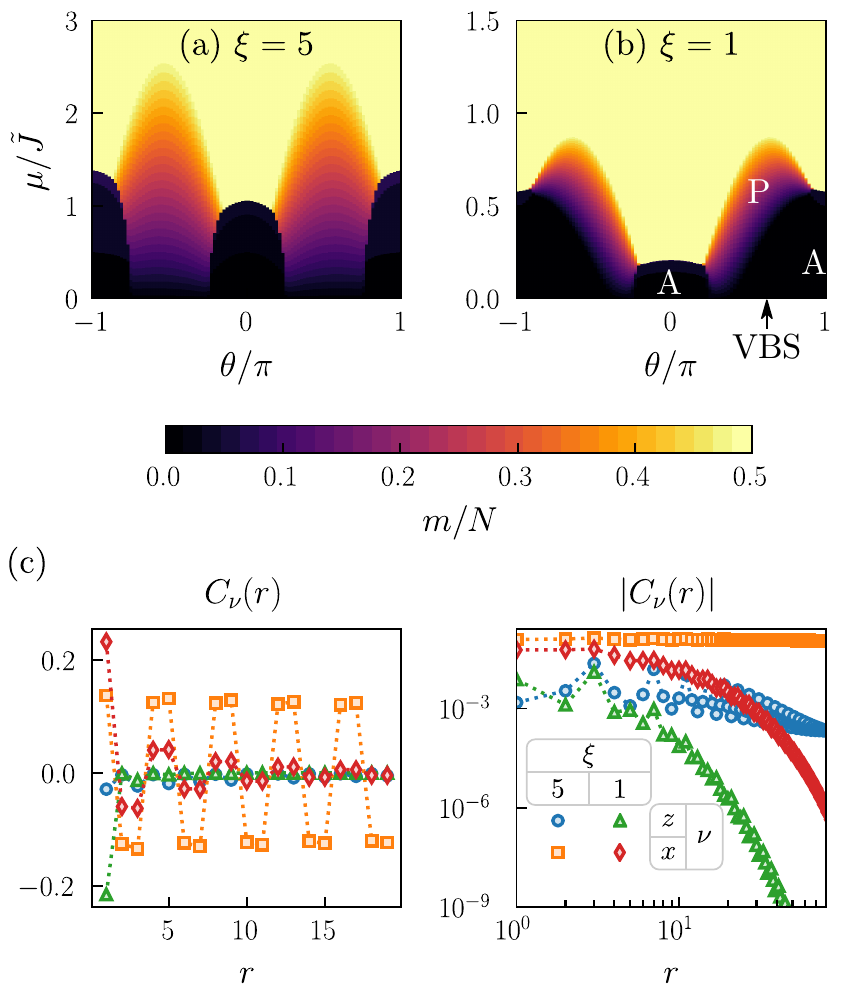}
  \caption{\textbf{Fully-dimerized ($\Delta=0$, $\deltaeff=\pm 1$), intermediate interaction-range, MBG model.} (a, b) Ground-state magnetization phase diagrams for chains with $N=60$ spins, for two different interaction lengths. (c) Correlations in the phase with zero magnetization $m=0$ for a chain with $N=120$ at $\theta=\pi/2$. \label{fig:fully_dimerized}}
\end{figure}

For $\Delta=0$ and $\delta\to 0^+$ [$\delta\to 0^-$] the only nonzero couplings are $J^{AB}_n=(-1)^{n}\Jeff$ [$J^{BA}_n=(-1)^{n-1}\Jeff$], in other words, the system is fully dimerized, $\deltaeff=\pm 1$. Note that in this case the system is always bipartite, even for a finite interaction lenght ($\xi > 0$). In chains with OBC, a negative value of the dimerization constant implies the presence of two uncoupled spins at the edges. Note that these edge modes are present regardless the value of the rest of parameters, that is, all phases for $\deltaeff < 0$ feature edge states---even when the bulk is in a gapless phase. These infinite-range models cannot be studied using collective spin operators, nonetheless, we can get an idea of its phase diagram looking at systems with large but finite $\xi$, see Fig.~\ref{fig:fully_dimerized}(a). For small interaction lengths we still observe VBS phases, planar and antiferromagnetic phases with a doubling of the period. Close to the classical limits we find gapped antiferromagnetic phases, where correlations along the z-axis are long-range ordered with a double N\'eel order, see Fig.~\ref{fig:fully_dimerized}(c). Between the two antiferromagnetic phases there are VBS phases for small values of $\mu$ and $\xi$. Again these VBS phases have a period consisting of two unit cells. When we increase the range of interactions, the VBS phase becomes less stable, eventually disappearing, leading to a gapless phase featuring a similar double N\'eel order as the antiferromagnetic phases but this time in the xy-plane.

\begin{figure}[tb]
  \includegraphics{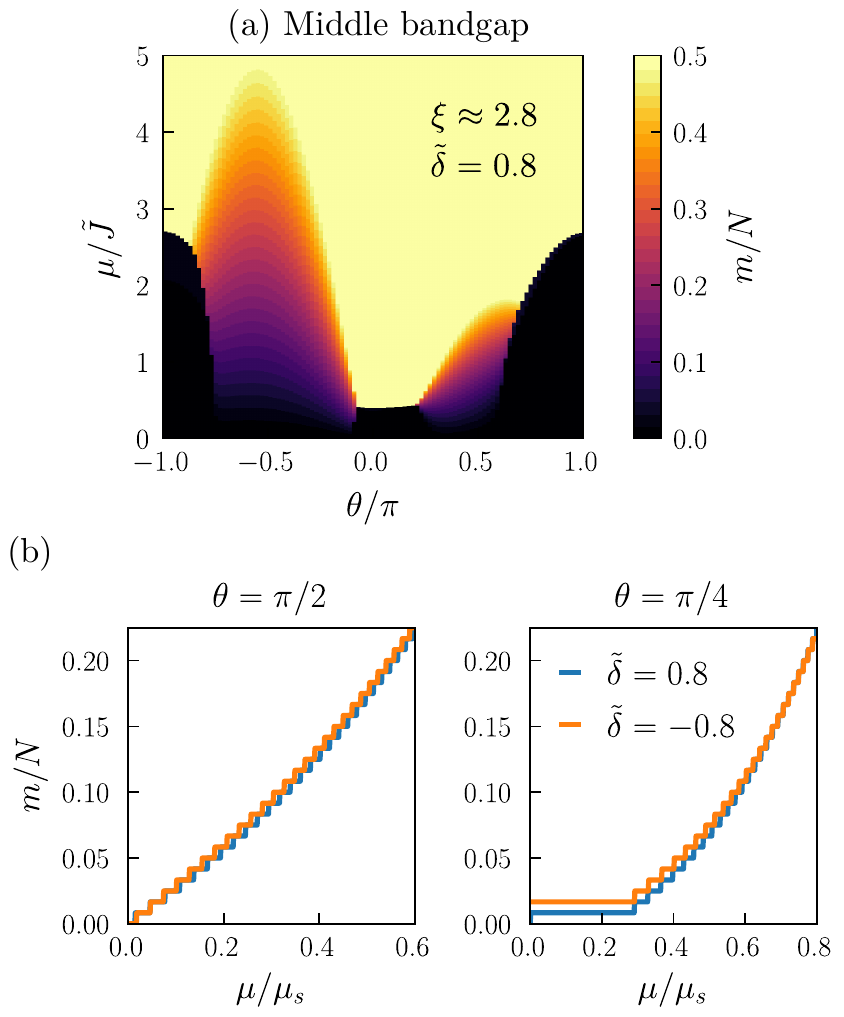}
  \caption{\textbf{Partially-dimerized ($\Delta\neq 0$, $\abs{\deltaeff} < 1$), intermediate interaction-range, MBG model.} (a) Ground state magnetization phase diagrams for a system with 
  $N=60$ spins in the MBG.
  (b) Magnetization curves for a longer chain with $N=120$ spins and the 
  same parameters.
  \label{fig:middle}}
\end{figure}

For systems in the MBG with $\Delta\neq 0$ the model is no longer 
bipartite, which makes the phase diagram asymmetric with respect to the 
change $\theta\to -\theta$, see Fig.~\ref{fig:middle}(a). We find phases
similar to the ones appearing for $\Delta = 0$. Surprisingly, the edge 
states of the gapless phases still survive when the ending spins are 
coupled to the rest of the chain, as can be seen in Fig.~\ref{fig:middle}(b)
comparing the magnetization curves of two open chains with 
$\deltaeff\gtrless 0$ and $\theta=\pi/4$.

\section{Adiabatic preparation~\label{sec:adiabatic}}


Another important aspect of analogue quantum simulators is how to reach the different many-body phases that can be explored within the simulators. An standard approach is to use adiabatic protocols~\cite{Albash2018}, in which one initially prepares the simulator in an easy-to-prepare state with the interactions switched off, and then turns on the interactions in such a way the ground-state of the systems follows adiabatically until the desired state. In this section we will precisely show how one could employ an adiabatic protocol to prepare one of the phases of interest of our model. For that, we assume a setup where interactions in the xy-plane and along the z-axis can be tuned independently, i.e., one can simulate Hamiltonians of the form $H=\tilde J(aH_z + bH_{xy})$, with $-1\leq a,b\leq 1$. Here,
\begin{align}
    H_z & = \frac{1}{\tilde J}\sum_{j,\,n > 0}\sum_{\alpha,\beta}J^{\alpha\beta}_n S^z_{j,\alpha}S^z_{j+n,\beta} \,, \\
    H_{xy} & = \frac{1}{\tilde J}\sum_{j,\,n > 0}\sum_{\alpha,\beta}\frac{J^{\alpha\beta}_n}{2}\left( S^+_{j,\alpha}S^-_{j+n,\beta} + \mathrm{H.c.}\right) \,,
\end{align}
are adimensional Hamiltonians, and $\tilde J$ denotes the largest coupling constant that can be produced in experiment. Additionally, we assume we can prepare any product state of up/down spins easily.

\begin{figure}[!htb]
    \centering
    \includegraphics{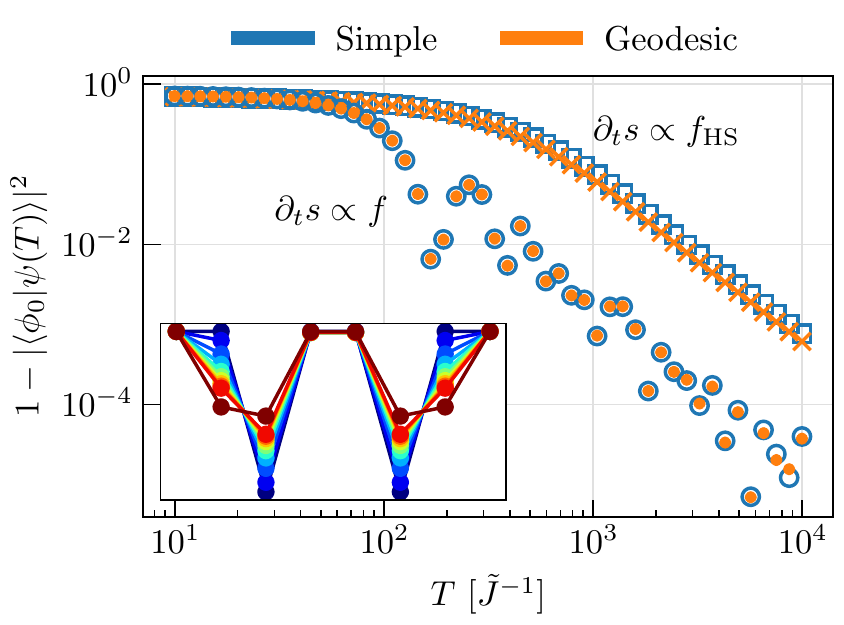}
    \caption{Infidelity of the final state $\ket{\psi(T)}$ with respect to the target ground state $\ket{\phi_0}$ as a function of the preparation time $T$ for different paths in parameter space and different schedule functions. The inset shows the evolution of the individual spin magnetization at different stages of the protocol (from blue in the beginning to red towards the end). The parameters of the model are: $N=8$, $N_\up = 6$, $\deltaeff = -0.2$, $\xi=2.0$; the emitters are tuned to the lower bandgap.}
    \label{fig:adiabatic_preparation}
\end{figure}

For the sake of concreteness, we will focus on preparing the ground state of the $\mathrm{VBS}_4$ phase in a system of $N=8$ emitters (spins). Note that in such small systems, there are no proper phase transitions, and we may only observe some of the features of the phases we are interested in. For example, in the case of the $\mathrm{VBS}_4$ phase, we would like to observe the formation of enlarged unit cells consisting of 4 spins, with the two middle spins forming a singlet, and the two outer spins polarized ``up'', see the inset in Fig.~\ref{fig:adiabatic_preparation}. 

Since any product state of up/down spins is an eigenstate of $H_z$, our starting Hamiltonian will be $H(0)\propto H_z$. Our target Hamiltonian is $H(T)\propto H_z + H_{xy}$ (Hamiltonian \eqref{eq:Hamiltonian} in the Heisenberg limit, $\theta = \pi/4$). To compare different protocols and simplify the discussion, we reparametrize the time-dependent Hamiltonian in terms of a (bijective) ``schedule function'' $s: [0, T] \to [0, 1]$, $H(t) = H(s(t))$. Ideally, we would like to employ an optimal modulation $\left(a(s), b(s)\right)$ that transforms the initial state into the target state within a given fidelity in the shortest amount of time. One possibility to design the protocol is to employ the geometrical approach developed in Ref.~\cite{Rezakhani2009}. They showed that for a schedule function satisfying $\partial_t s \propto \Delta^2(s)/\norm{\partial_s H}_\mathrm{HS}\equiv f_\mathrm{HS}(s)$, where $\Delta(s)$ is the energy gap between the ground and first-excited states of $H(s)$, and $\norm{\cdot}_\mathrm{HS}$ denotes the Hilbert-Schmidt norm, finding the optimal path in parameter space is equivalent to finding a shortest-distance geodesic in a Riemannian manifold, a problem that can be solved numerically using standard techniques \cite{Kimmel1998,Peyre2010}. In practice, for the problem at hand we find that a simple path of the form $a(s) = 1$, $b(s) = s$ is very close to the actual geodesic and yields very similar fidelities, see Fig.~\ref{fig:adiabatic_preparation}. Interestingly, we can improve the preparation time if we consider instead a schedule function satisfying $\partial_t s \propto \min_{n>0} \Delta^2_n(s)/\abs{\bra{\phi_n(s)}\partial_s H\ket{\phi_0(s)}}\equiv f(s)$, where $\Delta_n(s)$ denotes the energy gap between the instantaneous ground state $\ket{\phi_0(s)}$ and the $n$th instantaneous excited state $\ket{\phi_n(s)}$.

So far we have considered the fidelity of the target state only considering the imperfections introduced by the adiabatic protocols, in which the slower the process is made, the better fidelity is obtained (see Fig.~\ref{fig:adiabatic_preparation}). In real-life experiments, however, there are other decoherence sources affecting the fidelity of the final state. For concreteness, we will consider one of the simplest, but also more relevant, error sources that can occurr in waveguide QED setups, that is, the possibility of losing emitter excitations at rates $\gamma$. The evolution in presence of this kind of losses can be modelled through a Markovian master equation of the form 
\begin{equation}
    \partial_t \rho = -i[H(t), \rho] + \gamma \sum_j \mathcal{D}_j[\rho]\,,
\end{equation}
where $\mathcal{D}_j[\rho] \equiv S^-_j \rho S^+_j - \{S^+_jS^-_j, \rho\}/2$, and $\rho$ is the reduced density matrix of the emitters. For this dynamics, it can be shown that the fidelity with respect to the target ground state is $\mathcal{F}_\gamma = e^{-\gamma N_\up T} \mathcal{F}$, where $\mathcal{F}$ would be the fidelity if there were no losses, and $N_\up$ is the number of excitations in the target state. For example, for the simple-path protocol with $\partial_t s\propto f(s)$, we see that values of the decay rate $\gamma \lesssim 10^{-4}\tilde J$ allow to produce the target state with an 80\% fidelity or more.

We remark that the preparation time is limited essentially by the minimum gap of $H(s)$ throughout the protocol, which in the case analyzed occurs in the beginning, when $H(s)\simeq H_z$. The times shown in Fig.~\ref{fig:adiabatic_preparation} could be improved if one could enlarge the gap between the initial state and the rest of states. This can be done, for example, implementing local magnetic fields $\sum_j h_j S^z_j$, independently for each emitter with the appropriate pattern, and then turning them off towards the end of the protocol.


\section{Summary of the results~\label{sec:summary}}

Due to the wealth of different regimes and phases we cover along the manuscript, in this section we find it useful for the reader to summarize the main results in each situation. For the undimerized (standard) waveguide simulator there are two possible band-gap regions:
\begin{itemize}
    \item For the LBG, the spin interactions are long-range and with the same sign. This leads to physics qualitatively very similar to that found in other analog simulators with power-law interactions, e.g., trapped ions~\cite{hauke10c}, see Fig.~\ref{fig:undimerized_intermediate}(a). In particular, we observe a shrinking of the antiferromagnetic phase of the NN model, a \emph{Devil's stair case} of phases with different magnetizations around the classical limit ($\theta\approx 0$), and a spontaneously-dimerized phase at zero magnetization for $\theta\approx \pi/4$.
        
    \item For the UBG, the combination of the staggered-sign of the interactions with their long-range nature leads to a qualitatively different diagram, see Fig.~\ref{fig:undimerized_intermediate}(b). In particular, we find that the antiferromagnetic phase extends over a larger area of the phase diagram, and new gapless phases develop for $\theta\approx -3\pi/4$, such as the spin-density-wave and nematic phases that also appear in zig-zag Heisenberg chains~\cite{hikihara2008}.
\end{itemize}

For the topological (dimerized) waveguide, all the spin interactions inherit the dimerization of the underlying photonic bath. However, one can still find very different phases depending on which of the three different band-gaps the emitters are in resonance with:

\begin{itemize}

    \item In the LBG, the main effect of the dimerization is the appearance of new gapped phases, absent in the NN model, at certain non-zero magnetization values, see Fig.~\ref{fig:dimerized_lower}(a-b). They display large magnetic orderings, as corroborated by the calculation of the bond-order parameter $O_p$, see Fig.~\ref{fig:dimerized_lower}(c). We further understand these phases by connecting them to the equivalent fermionic model and studying the quantization of their many-body Berry phases~\cite{hatsugai2006,hirano2008,kariyado2018}, that enable us to further distinguish different SPT phases in them, depending on the sign of the dimerization constant. These new SPT phases are due to the unique dimerized long-range nature of the interactions appearing in such topological waveguide setups. 
        
    \item In the UBG regime, the dimerization affects the VBS phase around $\theta=-3\pi/4$ at zero magnetization, giving rise to different SPT phases depending on the sign of $\deltaeff$. It also affects the spin-density wave regime, as the system shows different magnetization steps as $\mu$ increases for $\tilde{\delta}\lessgtr 0$, which can be attributed to the appearance of persistent edge states.
        
    \item In the MBG regime for exactly $\Delta=0$ the spin interactions are fully dimerized. This has two consequences: i) it leads to the presence of uncoupled spins at the edges for the phases with $\tilde{\delta}<0$, no matter the the value of the rest of the parameters---even in the gapless phases; ii) the model displays VBS, planar, and antiferromagnetic phases reminiscent of the NN model, but with a double periodicity, see Fig.~\ref{fig:fully_dimerized}. When $\Delta\neq 0$, the interactions are not fully dimerized, but we find qualitatively similar phases as in the $\Delta=0$ case, including the appearance of edge states in the gapless phases (persistent edge modes).
        
\end{itemize}

\section{Conclusions~\label{sec:conclusions}}

To sum up, we have characterized the emergent many-body phases of a general class of spin models that can be simulated using quantum emitters subject to waveguide-mediated interactions. First, we have studied the effect of the tunable, long-range and possibly alternating-sign interactions that appear in standard waveguides. The interplay of all these features gives rise to phases different from those of the typical nearest-neighbor models, due to the larger frustration of the system. In addition, we have considered the impact of the dimerized interactions appearing in topological waveguides. The latter leads to the appearance of symmetry-protected topological phases, and, if they are also long-range, give rise to symmetry-protected topological phases with large magnetic orderings (larger than the period of the Hamiltonian). In all cases, we explain the experimental observables that can be measured to distinguish these phases based on either single-spin measurements (individual-spin magnetization) or two-spin correlations (bond-order parameter). Finally, we also show how the most interesting phases can be reached through adiabatic protocols, and analyze their fidelity in terms of the most typical error source. Overall, we believe our work uncovers the potential of waveguide QED setups as analogue quantum spin simulators, and can become the basis for future experiments on the subject. An interesting outlook would be to study the modification of these phase diagrams when the topological waveguides have larger winding numbers, since the shape of the resulting waveguide-mediated interactions can differ significantly \cite{vega2021}.

\begin{acknowledgements}
The authors thank Yasuhiro Hatsugai and Daniel Gonz\'alez-Cuadra for enlightening discussions. M. Bello acknowledges support from the ERC Advanced Grant QENOCOBA under the EU Horizon 2020 program (g.a.\ no.\ 742102), G. Platero acknowledges support from the Spanish Ministry of Science and Innovation through grant MAT2017-86717-P, and A. González-Tudela acknowledges support from the Spanish project PGC2018-094792-B-100(MCIU/AEI/FEDER, EU) and from the Proyecto Sinérgico CAM 2020 Y2020/TCS-6545 (NanoQuCo-CM). All authors acknowledge support from the CSIC Research Platform on Quantum Technologies PTI-001.  
\end{acknowledgements}

\appendix
\section{Coupling constants \label{sec:couplings}}

The collective self-energies have been obtained analytically in a recent work \cite{Bello2019}. For convenience, we rewrite them here. First, we define the following functions of the complex variable $z$, $r(z)=\sqrt{\prod_j(z-\omega_j)}$, with $\omega_j\in\{\pm 2J,\pm 2J\abs{\delta}\}$, and
\begin{equation}
  y_\pm(z)=\frac{z^2-2J^2(1+\delta^2)\pm r(z)}{2J^2(1-\delta^2)}\,.
\end{equation}
The collective self energies $\Sigma^{\alpha\beta}_{m,n+m}\equiv \Sigma^{\alpha\beta}_n$ read
\begin{align}
  \Sigma^{AA}_n(z) & = \mp\frac{g^2z{y_\pm}^{\abs{n}}}{r(z)}\,,\\
  \Sigma^{AB}_n(z) & = \pm\frac{g^2 J\left[(1+\delta){y_\pm}^{\abs{n}}
  +(1-\delta){y_\pm}^{\abs{n+1}}\right]}{r(z)}\,,
\end{align}
where the upper (lower) sign has to be chosen if $\abs{\re(z)}<2J\abs{\delta}$ ($\abs{\re(z)}>2J$). Furthermore, $\Sigma^{BB}_n(z)=\Sigma^{AA}_n(z)$, and $\Sigma^{BA}_n(z)=\Sigma^{AB}_{-n}(z)$ for $n\geq 1$. In these expressions $g$ denotes the light-matter coupling constant, i.e., the coupling constant determining the strength of the interaction between the emitter and the local fields in the waveguide.

\begin{figure}[!htb]
  \includegraphics{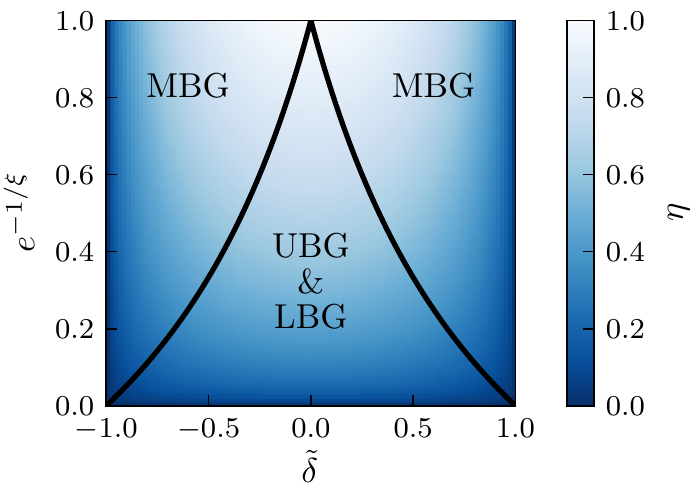}
  \caption{Allowed values for the parameters ($\xi,\deltaeff,\eta$) that 
  determine the form of the interactions $J^{\alpha\beta}_n$. The black 
  lines delimit the parameter regions valid for the upper and lower bandgaps
  (UBG \& LBG), and the middle bandgap (MBG).
  \label{fig:allowedparam}}
\end{figure}

As explained in the main text, the spin-spin interaction constants in the Markovian regime are determined by $J^{\alpha\beta}_n = \re[\Sigma^{\alpha\beta}_n(\Delta+i0^+)]$. We can rewrite them in a more convenient form, see Eqs.~(\ref{eq:JAB}--\ref{eq:JAA}), defining an interaction length as $\xi = -1/\log\abs{y_\pm}$; an effective dimerization constant 
\begin{equation}
  \begin{split}
  \deltaeff & \equiv\frac{\abs{J^{AB}_{n}} - \abs{J^{BA}_{n + 1}}}
  {\abs{J^{AB}_{n}} + \abs{J^{BA}_{n + 1}}} \\
    & = \frac{1-e^{-1/\xi}}{1+e^{-1/\xi}}\times
  \begin{cases}
    \delta^{-1} & \text{if }\abs{\Delta}<2J\abs{\delta} \\
    \delta & \text{if }\abs{\Delta}>2J 
  \end{cases} \,,\\
  \end{split}\label{eq:deltaeff}
\end{equation}
which corresponds to the dimerization of the waveguide-mediated spin interactions; and the ratio 
\begin{equation}
  \begin{split}
    \eta & \equiv\frac{2\abs{J^{AA}_{n+1}}}{\abs{J^{AB}_n} 
    + \abs{J^{BA}_{n+1}}}\\
    & =\frac{\abs{\Delta}e^{-1/\xi}}{J(1+e^{-1/\xi})}\times
  \begin{cases}
    \abs{\delta}^{-1} & \text{if }\abs{\Delta}<2J\abs{\delta} \\
    1 & \text{if }\abs{\Delta}>2J
  \end{cases}\,.\\
  \end{split}
\end{equation}
Note that $\deltaeff$ and $\delta$ have both the same sign, so the dimerization pattern of the effective spin model follows that of the underlying bath. Also, form Eq.~\eqref{eq:deltaeff}, we can see that the pairs $(\xi,\deltaeff)$ such that $\abs{\deltaeff}<(1-e^{-1/\xi})/(1+e^{-1/\xi})$ are possible only in the outer bandgaps, while the pairs satisfying $\abs{\deltaeff}>(1-e^{-1/\xi})/(1+e^{-1/\xi})$ are possible only in the middle bandgap. Fig.~\ref{fig:allowedparam} summarizes at a glance the possible values of the parameters for the different models.

The global strength of the spin interactions can be computed as 
\begin{equation}
  \Jeff=\frac{g^2J(1+e^{-1/\xi})}{r(\Delta)}\times
  \begin{cases}
    \abs{\delta} & \text{if }\abs{\Delta}<2J\abs{\delta}\\
    1 & \text{if }\abs{\Delta}>2J 
  \end{cases} \,.
\end{equation}

\section{Infinite-range Hamiltonian in the UBG and LBG \label{sec:IRupper}}

The action of the Hamiltonian in Eq~\eqref{eq:infinite_range_upper} on the states $\ket{s_A,m_A;s_B,m_B}$, which are eigenstates of ${\mathbf S}^2_\alpha$ and $\s{z}{\alpha}$ with eigenvalues $s_\alpha(s_\alpha + 1)$ and $m_\alpha$ respectively ($\alpha=A,B$), is

\begin{widetext}
  \begin{multline}
    H\ket{s_A,m_A;s_B,m_B}=\\-\tilde{J}
    \left\{\frac{\sin\theta}{2}\left[s_A(s_A+1) - m^2_A + s_B(s_B+1) - m^2_B\right]
    + \frac{\cos\theta}{2}(m_A - m_B)^2 + \frac{\mu}{\tilde J}(m_A + m_B)\right\}
    \ket{s_A,m_A;s_B,m_B}\\
    + \tilde{J}\frac{\sin\theta}{2}\sqrt{s_A(s_A + 1) - m_A(m_A + 1)}
    \sqrt{s_B(s_B + 1) - m_B(m_B - 1)}\ket{s_A,m_A+1;s_B,m_B-1}\\
    + \tilde{J}\frac{\sin\theta}{2}\sqrt{s_A(s_A + 1) - m_A(m_A - 1)}
    \sqrt{s_B(s_B + 1) - m_B(m_B + 1)}\ket{s_A,m_A-1;s_B,m_B+1}\,.
  \end{multline}
\end{widetext} 

Thus, the Hamiltonian can be represented by a tridiagonal matrix when acting on the subspace of states with fixed $(s_A, s_B, m=m_A+m_B)$, if a properly sorted basis is chosen.
We have neglected the permutation quantum numbers, since they don't play any role, but the degeneracy of the ground-state manifold is directly related to them. Numerically, looking for the ground state(s) within each sector with definite total magnetization $m\geq 0$, we find that for $-\pi<\theta<-\pi/2$ there is a  ground state with total angular momenta $s_A=s_B=\ceil{m/2}$ ($\ceil{x}$ denotes the closest integer or half-integer larger than $x$, depending on whether $N/2$ is even or odd). For $-\pi/2<\theta<0$, there is a ground state such that $s_A=N/4$ and $s_B=\abs{m-N/4}$, or vice versa. For $0<\theta<\pi$ the ground state has maximum total angular momenta $s_A=s_B=N/4$. As it can be seen in Fig.~\ref{fig:infinite_range}(b) for $-\pi/2<\theta<0$ the ground state either has zero total magnetization or is the fully polarized state. So, for $-\pi/2<\theta<\pi$ the ground state is unique and is fully symmetric under permutations of spins within each sublattice. This symmetry entails homogeneous correlations, which do not depend on distance, but are different for spins within the same/opposite sublattice. They can be computed easily once we have obtained the ground state, as
\begin{equation}
    \mean{S^\nu_{i,\alpha}S^\nu_{j,\beta}} = \begin{cases}
    \frac{4}{N^2}\mean{S^\nu_A S^\nu_B} & \text{if } \alpha\neq\beta \\
    \frac{4}{N(N-2)}\mean{S^\nu_\alpha S^\nu_\alpha} - \frac{1}{2N-4} & \text{if }\alpha = \beta
    \end{cases}\,,
\end{equation}
where $S^\nu_\alpha$ denote the collective spin operators already employed to rewrite the Hamiltonian [Eq.~\eqref{eq:infinite_range_upper}], which have again a simple, at most tridiagonal, matrix representation in the relevant subspace.

For $-\pi<\theta<-\pi/2$ the ground state magnetization takes the values $m\in\{0,N/2\}$ ($m\in\{0,1,N/2\}$) if $N/2$ is even (odd), see Fig.~\ref{fig:infinite_range}(d). In these cases the relevant subspaces have a very small dimension, such that the ground states can be found analytically, and with them also the correlations. If $N/2$ is even, the correlations in the phase with $m=0$ are $\mean{S^\nu_{i,\alpha}S^\nu_{j,\alpha}}=-1/(2N-4)$, $\mean{S^\nu_{i,A}S^\nu_{j,B}}=0$. If $N/2$ is odd, then for $m=0$, $\mean{S^\nu_{i,\alpha}S^\nu_{j,\alpha}}=-\frac{1}{2N}$, $\mean{S^z_{i,A}S^z_{j,B}}=-1/N^2$ and $\mean{S^x_{i,A}S^x_{j,B}}=1/N^2$; whereas for $m=1$ $\mean{S^\nu_{i,\alpha}S^\nu_{j,\alpha}}=-\frac{1}{2N}$, $\mean{S^z_{i,A}S^z_{j,B}}=1/N^2$ and $\mean{S^x_{i,A}S^x_{j,B}}=0$.

\section{Nematic and spin-density-wave phases \label{sec:nematic}}

Our model in the upper bandgap with $\deltaeff = 0$ and $\theta = -3\pi/4$
resembles the zigzag Heisenberg model with ferromagnetic NN and 
antiferromagnetic NNN interactions studied in Ref.~\cite{hikihara2008}. 
However, our model has longer range interactions as well, and we cannot 
increase the ratio of NNN to NN interactions without increasing them too. 
Among all the phases appearing in the zigzag Heisenberg model, the nematic 
phase and a spin-density wave ($\mathrm{SDW}_2$) phase also appear in our 
model. This is demonstrated in Fig.~\ref{fig:nematic} below, where we show 
several correlation functions that characterize the different phases,
\begin{align}
  C_2(r) & = 
  \mean{\s{+}{n-1}\s{+}{n}\s{-}{n+r}\s{-}{n+r+1}} \,, \\
  C_3(r) & = \mean{\s{+}{n-2}\s{+}{n-1}\s{+}{n}
  \s{-}{n+r}\s{-}{n+r+1}\s{-}{n+r+2}} \,,\\
  C_\kappa(r) & = \mean{\kappa_{n}\kappa_{n + r}} \,,
\end{align}
where $\kappa_n=\left(\mathbf{S}_n\times\mathbf{S}_{n + 1}\right)^z=
\s{x}{n}\s{y}{n + 1} - \s{y}{n}\s{x}{n + 1}$.

\begin{figure}[!htb]
  \includegraphics{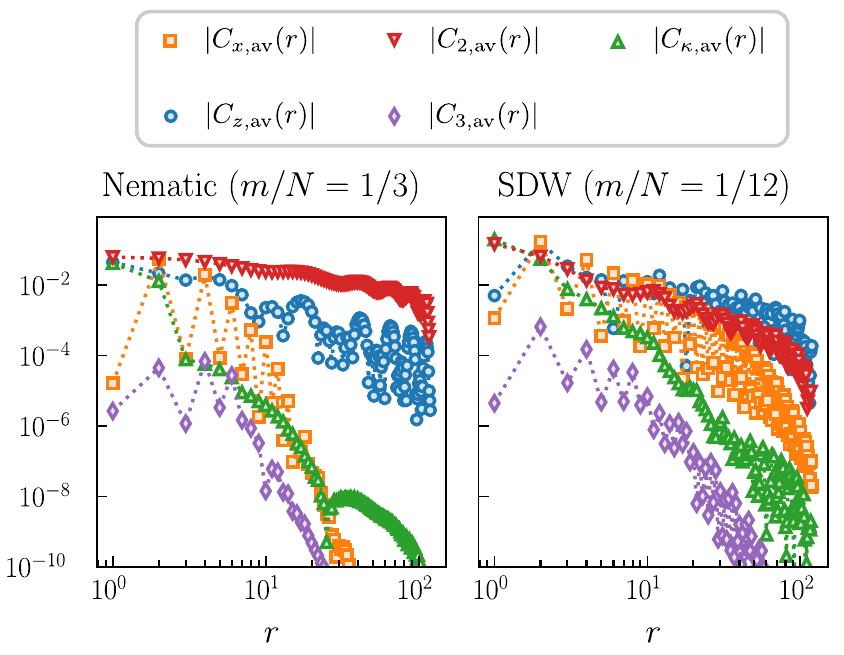}
  \caption{Different correlation functions for the undimerized model 
  ($\deltaeff = 0$) with $\xi=3$, $\theta=-3\pi/4$, in the upper bandgap. 
  The correlations have been computed for a chain of length $N=120$, taking 
  the average over the pairs of spins that are closest to the middle of 
  the chain. \label{fig:nematic}}
\end{figure}

\section{Dimerized XX Model \label{sec:dimerXX}}

We would like to solve for the ground state of the following 
Hamiltonian 
\begin{equation}
  \begin{split}
    H & = \frac{\Jeff}{2}\sum_j\left[1+(-1)^j\deltaeff\right]
          \left(\s{+}{j}\s{-}{j+1}+\s{-}{j}\s{+}{j+1}\right)\\
      &\quad -\mu\sum_j\s{z}{j}\,.
  \end{split}
\end{equation}
As in the main text, we will go back and forth between the
single-index and two-indices labelling of the different spins.
Following \cite{Lieb1961}, we proceed by applying the Jordan-Wigner 
transformation:
\begin{equation}
  \s{-}{j} = c_j\prod_{n<j}(1-2c^\dagger_nc_n)\,,\
  \s{z}{j} = c^\dagger_j c_j - 1/2\,.
\end{equation}
The Hamiltonian becomes
\begin{equation}
  \begin{split}
    H & = \frac{\Jeff}{2}\sum_j\left[1+(-1)^j\deltaeff\right]
    \left(c^\dagger_{j}c_{j+1}+\mathrm{H.c.}\right)\\
    &\quad-\mu\sum_j\left(c^\dagger_jc_j - \frac{1}{2}\right)\,,
  \end{split}
\end{equation}
which corresponds to a model of free fermions in a 1D lattice. Actually,
it is exactly the standard SSH model with a Fermi level set by $\mu$. 
Calling $a_j\equiv c_{2j}$ ($b_j\equiv c_{2j+1}$), we can diagonalize it
by Fourier transforming $a_j=(N/2)^{-1/2}\sum_k e^{ikj}a_k$ 
[$b_j=(N/2)^{-1/2}\sum_k e^{ikj}b_k$] (remember that $N$ denotes the number 
of emitters, which is twice the number of unit cells). Its energy spectrum 
consists of two bands
\begin{equation}
  H=\sum_k \left(\epsilon^u_k u^\dagger_ku_k + 
  \epsilon^l_k l^\dagger_kl_k\right)\,,
\end{equation}
with dispersion relations and single-particle eigenstates given by:
\begin{align}
  \epsilon^{u/l}_k & = 
  \pm\frac{\Jeff}{2}\sqrt{2(1+\delta^2) + 2(1 - \delta^2)\cos(k)}
  -\mu\,,\\
  u_k/l_k & =\left(\pm a_k+e^{i\phi_k}b_k\right)/\sqrt{2}\,,
\end{align}
with $\phi_k=\arg\left((1+\delta)+(1-\delta)e^{-ik}\right)$. The many-body 
eigenstate with lowest energy is the one in which all single-particle 
levels with negative energies are occupied. For $\mu\geq 0$ this implies 
that excitations over the ground state are created by the following 
fermionic creation operators
\begin{equation}
  \eta^\dagger_k=l_k\,,\ 
  \xi^\dagger_k=\begin{cases}u^\dagger_k\,, & \abs{k}\leq k_0\\
    u_k\,, & \abs{k}>k_0
  \end{cases}\,.
\end{equation}
Here $k_0$ denotes the quasimomentum of the Fermi points, i.e., $k_0>0$ such
that $\epsilon^u_{k_0}=0$. This equation has solution only when 
$\Jeff\abs{\delta}\leq\mu\leq \Jeff$. Outside of this range, we also 
define $k_0$ as follows for convenience:
\begin{equation}
  k_0=\begin{cases}\pi\,, & 0\leq\mu\leq\Jeff\abs{\deltaeff}\\ 
    \arccos\left(\frac{2\mu^2/\Jeff^2 - 1 - \deltaeff^2}{1 - \deltaeff^2}\right)\,,
    & \Jeff\abs{\deltaeff}<\mu<\Jeff\\
    0\,, & \Jeff\leq\mu\\
  \end{cases}\,.
\end{equation}                             
The ground state $\ket{\mathrm{GS}}$ satisfies $\xi_k\ket{\mathrm{GS}}
=\eta_k\ket{\mathrm{GS}}=0$. In the reminder of this section, all 
expectation values are taken with respect to this ground state. The 
ground state magnetization is 
\begin{equation}
  m=\frac{1}{N}\sum_j\langle a^\dagger_j a_j + b^\dagger_j b_j - 1\rangle
  =\frac{\pi-k_0}{2\pi}\,.
\end{equation}

We can compute correlations between any two spins in the chain as follows
\begin{align}
  \mean{\s{z}{i}\s{z}{j}}&=\frac{1}{4}\mean{A_iB_iA_jB_j}\,,\\
  \mean{\s{x}{i}\s{x}{j}}&=\frac{1}{4}\mean{B_i
  \left(\prod_{n=i+1}^{j-1}A_nB_n\right)A_j}\,,
\end{align}
with
\begin{equation}
  A_i = c^\dagger_i + c_i\,, \ B_i = c^\dagger_i - c_i \,.
\end{equation}
This is valid for any 1D spin chain. In our case, we define 
$A^a_i/B^a_i \equiv A_{2i}/B_{2i} = a^\dagger_i \pm a_i $, and 
$A^b_i/B^b_i \equiv A_{2i+1}/B_{2i+1} = b^\dagger_i\pm b_i$. The 
contractions of these operators are 
$\mean{A^\alpha_iA^\beta_j}=\delta_{\alpha\beta}\delta_{ij}$,
$\mean{B^\alpha_iB^\beta_j}=-\delta_{\alpha\beta}\delta_{ij}$,
$\mean{B^\alpha_iA^\alpha_j}=G_{i-j}$, $\mean{B^a_iA^b_j}=F_{i-j}$, and
$\mean{B^b_iA^a_j}=F_{j-i}$, with
\begin{align}
  G_r & \equiv \frac{1}{\pi}\int_{k_0}^\pi dk\,\cos(kr)=
  \begin{cases}1-k_0/\pi\,, & r=0\\
  -\frac{\sin(k_0r)}{\pi r}\,, & r\neq 0\end{cases}\,,\\
  F_r & \equiv-\frac{1}{\pi}\int_0^{k_0}dk\,\cos(kr+\phi_k)\,.
\end{align}
Using Wick's theorem we can now compute
\begin{align}
  \mean{\s{z}{i}\s{z}{j}} & = \frac{\mean{B_iA_i}\mean{B_jA_j}-
  \mean{B_iA_j}\mean{B_jA_j}}{4}\,,\\
  \mean{\s{x}{i}\s{x}{j}} & = \frac{\det M}{4}\,,
\end{align}
with 
\begin{equation}
  M = \begin{pmatrix}
  \mean{B_iA_{i+1}} & \mean{B_iA_{i+2}} & \cdots & \mean{B_iA_{j}} \\
  \mean{B_{i+1}A_{i+1}} & \mean{B_{i+1}A_{i+2}} & \cdots & \mean{B_{i+1}A_j}\\
  \vdots & \vdots & \ddots & \\
  \mean{B_{j-1}A_{i+1}} & \mean{B_{j-1}A_{i+2}} & & \mean{B_{j-1}A_j}
  \end{pmatrix}\,.
\end{equation}
The dimer order parameter, Eq.~\eqref{eq:bondorder} particularized for 
$p = 2$, can be computed as
\begin{equation}
\begin{split}
  O_2 & =\frac{\mean{\vs{n,B}\vs{n+1,A}}- \mean{\vs{n,A}\vs{n,B}}}{2}\\
  & = \frac{F_0^2-F_1^2+2(F_1-F_0)}{8}\,.
\end{split}
\end{equation}


\section{non-Abelian, many-body Berry phases \label{sec:berry}}

As discussed in Refs.~\cite{hatsugai2006,kariyado2014,kariyado2018}, the many-body Berry phases obtained for certain specific twists of the Hamiltonian can be interpreted as topological invariants characterizing different SPT orders. For this, one has to choose an appropriate modulation of the Hamiltonian, such that the associated Berry phases are quantized due to the symmetries of the phase we want to characterize. 

Let us apply this theory to the $\mathrm{VBS}_q$ phases discussed in the main text. As we will show, a suitable twist to characterize these phases consist in the modulation of the complex phase of the exchange coupling constant between two spins in \emph{different sublattices}: $e^{-i\phi}S^+_{m, A} S^-_{n, B} + \mathrm{H.c.}$. If the ground state is unique and it is symmetric about the middle of the link connecting these two spins, then the Berry phase 
\begin{equation}
    \gamma = -\int_0^{2\pi} d\phi\, \bra{\mathrm{GS}(\phi)} i\partial_\phi \ket{\mathrm{GS}(\phi)}\,,
\end{equation}
obtained for the ground state $\ket{\mathrm{GS}(\phi)}$ of the modulated Hamiltonian $H(\phi)$, must be quantized. To prove it, let us consider also the complementary path in which $\phi$ goes from $2\pi$ to 0, and the associated Berry phase $\gamma'$. Since the concatenation of both paths yields the trivial path, we have that $\gamma + \gamma' = 0$ (mod $2\pi$). On the other hand, for chains with PBC, $H(-\phi) = U_I H(\phi) U^\dagger_I$, where $U_I$ is the (non-local) unitary associated to the space-inversion transformation about the middle of the link connecting the two spins. If for all values of $\phi$ the system remains gaped, and the ground state is unique, the ground states along each path are related to each other through a gauge transformation, $\ket{\mathrm{GS}(-\phi)} = U_I \ket{\mathrm{GS}(\phi)} = e^{i\Omega(\phi)}\ket{\mathrm{GS}(\phi)}$, which, in turn, implies that $\gamma = \gamma'$ (mod $2\pi$). These two conditions combined restrict the possible values of $\gamma$ to $0$ or $\pi$ (mod $2\pi$).

A similar derivation can be obtained for non-Abelian Berry phases in the case of a degenerate ground state \cite{wilczek1984}. This is the one we will use to characterize the ground state in the $\mathrm{VBS}_q$ phases, for $q > 2$. Numerically, it can be computed discretizing the path, choosing a set of points $\{\phi_n\}_{n=1,\dots,N}$ such that $0\leq \phi_n < \phi_{n+1}< 2\pi$, and computing the matrix of overlaps $\Phi_n$, whose components are $(\Phi_n)_{\mu\nu} = \braket{\mathrm{GS}_\nu(\phi_n)\vert\mathrm{GS}_\mu(\phi_{n+1})}$, where $\{\ket{\mathrm{GS}_\mu(\phi_n)}\}_{\mu=1,\dots,q/2}$ denotes the ground state multiplet at each point along the path (we identify $\ket{\mathrm{GS}_\mu(\phi_{N+1})} = \ket{\mathrm{GS}_\mu(\phi_{1})}$). Then, the non-Abelian Berry phase can be computed as $\gamma = -\sum_n \arg(\det \Phi_n)$. Even though a single many-body ground state may be periodic with a period larger than the period of the Hamiltonian, when we compute these non-Abelian Berry phases, we recover the original periodicity of the Hamiltonian, so that there are just two inequivalent many-body Berry phases between nearest-neighbor spins in chains with PBC: $\gamma_\mathrm{inter}$ and $\gamma_\mathrm{intra}$. Furthermore, since changing the sign of the dimerization parameter is equivalent to a redefinition of the unit cell, it is clear that upon chaging $\deltaeff \to -\deltaeff$ the two Berry phases interchange $\gamma_\mathrm{inter}\leftrightarrow\gamma_\mathrm{intra}$.

\bibliography{alexrefs,myrefs,AdiabaticPreparation/adiabatic_preparation}

\end{document}